\documentclass[prl,aps,reprint,twocolumn]{revtex4}
\usepackage{graphics,graphicx}
\usepackage{color}
\usepackage{wrapfig}
\usepackage{enumerate}
\usepackage{amssymb,amsmath}
\usepackage{bm}
\usepackage{mhchem}

\begin{document}
\title{Hysteretic magnetoresistance and unconventional anomalous Hall effect in the frustrated magnet \ce{TmB4}}
\author{Sai~Swaroop~Sunku,$^1$  Tai~Kong,$^2$ Toshimitsu~Ito,$^3$ Paul~C.~Canfield,$^2$ B.~Sriram~Shastry,$^4$ Pinaki~Sengupta,$^1$ and Christos~Panagopoulos$^1$}
\affiliation{$^1$Division of Physics and Applied Physics, School of Physical and Mathematical Sciences, Nanyang Technological University, 21 Nanyang Link, Singapore 637371}
\affiliation{$^2$Ames Laboratory, U.S. DOE and Department of Physics and Astronomy, Iowa State University, Ames, Iowa 50011, USA}
\affiliation{$^3$National Institute of Advanced Industrial Science and Technology (AIST), Tsukuba, Ibaraki 305-8562, Japan}
\affiliation{$^4$Physics Department, University of California, Santa Cruz, CA 95064, USA}
\date{\today}

\begin{abstract}
We study \ce{TmB4}, a frustrated magnet on the Archimedean Shastry-Sutherland lattice, through magnetization and transport experiments. The lack of anisotropy in resistivity shows that \ce{TmB4} is an electronically three-dimensional system. The magnetoresistance (MR) is hysteretic at low-temperature even though a corresponding hysteresis in magnetization is absent. The Hall resistivity shows unconventional anomalous Hall effect (AHE) and is linear above saturation despite a large MR. We propose that complex structures at magnetic domain walls may be responsible for the hysteretic MR and may also lead to the AHE.
\end{abstract}

\maketitle

Geometric frustration in magnetic systems arises from competing magnetic interactions that cannot be satisfied simultaneously and leads to a variety of exotic ground states \cite{Lacroix2011}. While insulating frustrated materials are well studied, metallic systems have received less attention \cite{Julian2012}. In metallic materials, the conduction electrons mediate interactions between the magnetic moments. Additionally, the transport properties in such systems can be strongly influenced by the magnetic structure \cite{Lacroix2011}. This interplay between magnetism and charge can be exploited in two ways: to engineer a highly field tunable response of the transport properties \cite{Ueland2012} or to use transport experiments as an indirect probe of the complex magnetic structures that arise in such systems \cite{Taguchi2001, Machida2010}.

\begin{figure}[!b]
	\includegraphics[width=\linewidth]{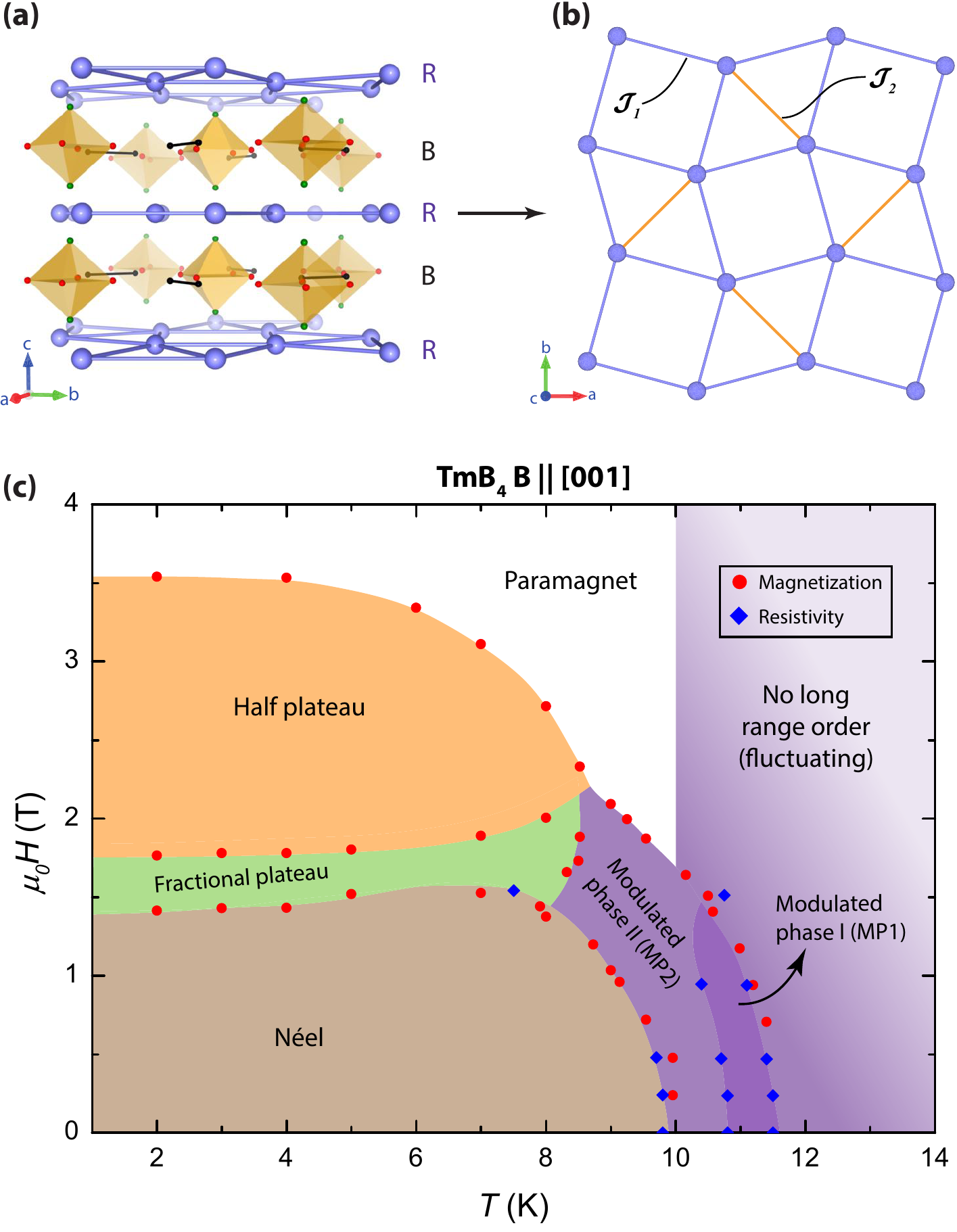}
	\caption{\textbf{(a)} Crystal structure of \ce{RB4}. The R and B layers are labelled. \textbf{(b)} The R sublattice viewed along the c-axis, showing the {Archimedean Shastry-Sutherland lattice}. {\textbf{(c)} Phase diagram of \ce{TmB4} as determined from our data \cite{Suppl}.}}
\label{Fig_phase_diag}
\end{figure}

The rare earth tetraboride family (\ce{RB4}, R is a rare earth) is a series of metallic frustrated magnets. \ce{RB4} crystallizes in a tetragonal structure (space group {\em P4/mbm}, 127) \cite{Fisk1972}, consisting of alternating layers of R and B ions (Fig \ref{Fig_phase_diag}(a)). The \ce{R} ions form a frustrated Shastry-Sutherland lattice (SSL) with competing interactions $\mathcal{J}_1$ and $\mathcal{J}_2$ \cite{Shastry1981}. Quite remarkably, high resolution structural refinement of \ce{LaB4} \cite{Kato1974} and \ce{HoB4} \cite{Olsen2011} show that the R-R bonds corresponding to $\mathcal{J}_1$ and $\mathcal{J}_2$ are equal in length, making the R-sublattice a rare physical realization of one of the eleven Archimedean lattices \cite{Farnell2014} (Fig. 1(b)). While other frustrated Archimedean lattices such as the triangular and Kagom\'{e} lattices are well studied \cite{Harrison2004, Farnell2014}, the \ce{RB4} family is the only known realization of the {Archimedean Shastry-Sutherland lattice}.

{In this article, we use magnetization and transport experiments to study \ce{TmB4}}, a member of the \ce{RB4} family that has attracted attention for its rich phase diagram \cite{Iga2007, Siemensmeyer2008, Gabani2008, Michimura2009} (Fig. \ref{Fig_phase_diag}(c)). Crystal field effects at the \ce{Tm^{3+}} sites (site symmetry $mm$) lift the degeneracy of the $J = 6$ multiplet and the ground state is the doublet $M_J = \pm6$ \cite{Siemensmeyer2008}. A strong Ising anisotropy is present \cite{Yoshii2006} and the interactions between the \ce{Tm^{3+}} spins consist of both direct exchange and  RKKY. Below $T_{N2} = 9.7$K, an antiferromagnetic N\'{e}el phase is stable and the magnetization shows a striking field dependence: a wide half plateau is present at $M/M_{sat} = 1/2$ ($M_{sat}$ is the saturation magnetization of 7$\mu_B$/Tm) and a narrow hysteretic fractional plateau at $M/M_{sat} \sim 1/8$ \cite{Siemensmeyer2008, Gabani2008, Wierschem2015}. Between $T_{N1}$ = 11.7K and $T_{N2}$, neutron scattering experiments find two long-range-modulated phases, MP1 and MP2 \cite{Michimura2009}. While MP1 can be indexed by a single modulation vector of periodicity $\sim$ 8 unit cells (u.c.), MP2 requires an additional modulation of $\sim$ 80 u.c. \cite{Michimura2009}. Frustration in \ce{TmB4} is reflected in the moderately large frustration parameter \cite{Gabani2008, Kim2013} and in the appearence of a diffuse peak in neutron scattering above $T_{N2}$ \cite{Michimura2009}, indicative of short-range order. In the temperature range $T_{N1} > T > T_{N2}$, the diffuse peak coexists with the sharp peaks from MP1 and MP2 \cite{Michimura2009}.

Theoretical models for \ce{TmB4}, focused on explaining the unusual plateau structure, have assumed a two dimensional (2D) nature (in analogy to another SSL compound \ce{SrCu(BO3)2} \cite{Sebastian2008}). While a 2D SSL in the Ising limit cannot have a half plateau \cite{Liu2009}, several groups have demonstrated the existence of a half plateau by considering longer range interactions \cite{Chang2009, Suzuki2009, Suzuki2010, Dublenych2012}. Even so, the modulated phases and the fractional plateau remain unexplained, despite the relatively simple structure of \ce{TmB4} and intense theoretical effort  \cite{Liu2009, Chang2009, Suzuki2009, Suzuki2010, Dublenych2012}.

{Here we present a combined transport and magnetization {study} of \ce{TmB4}.} By measuring the resistivity anisotropy, we find that \ce{TmB4} is an electronically three dimensional (3D) system. We find unusual hysteretic magnetoresistance (MR) which {may} arise from complex structures at magnetic domain walls. We further find the presence of an unconventional anomalous Hall effect (AHE).

\textit{Methods} - TmB$_4$ single crystals were synthesized by solution growth method using an Al flux and oriented using X-ray diffraction in the Laue geometry to within $\pm 5^{\circ}$ \cite{Suppl}. Quantum Design (QD) MPMS XL SQUID magnetometer was used for DC magnetization measurements and QD PPMS for transport experiments \cite{Suppl}. Since the magnetization in the fractional plateau phase is known to vary with field history \cite{Siemensmeyer2008,Wierschem2015}, a protocol was developed that reproduces the same magnetization curve at 2K when the measurement is repeated  \cite{Suppl}.

\begin{figure}[!]
	\includegraphics[clip,trim=0cm 0cm 0cm 0cm, width=\linewidth]{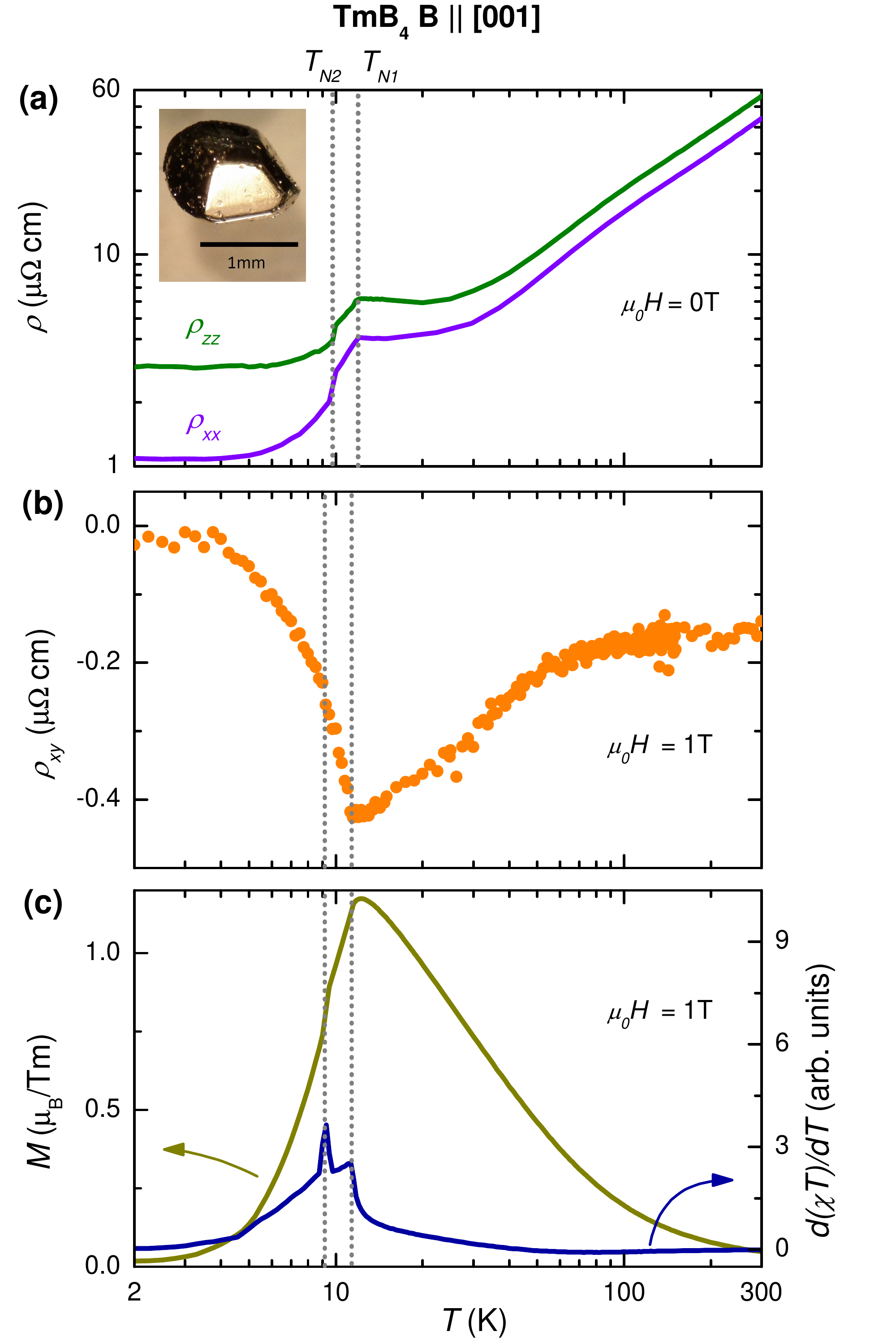}
	\caption{\textbf{(a)} In plane ($\rho_{xx}$) and out of plane ($\rho_{zz}$) longitudinal resistivities at zero field. Inset: photograph of a \ce{TmB4} single crystal used in our experiments. The c-axis is perpendicular to the shiny facet. \textbf{(b)} In plane Hall resistivity $\rho_{xy}$ at $\mu_0H = 1$T. \textbf{(c)} Magnetization $M$ and $d(\chi T)/dT$ ($\chi$ is the dc susceptibility) at $\mu_0H = 1$T. The magnetic field is not corrected for demagnetization. Vertical dotted lines represent $T_{N1}$ and  $T_{N2}$. We estimate an error of 20\% on the absolute values of $\rho_{xx}$, $\rho_{xy}$ and $\rho_{zz}$ \cite{TransportError}.}
\label{Fig_temp}
\end{figure}

\begin{figure*}[t]
	\includegraphics[clip,trim=0cm 0cm 0cm 0cm, width=17.2cm]{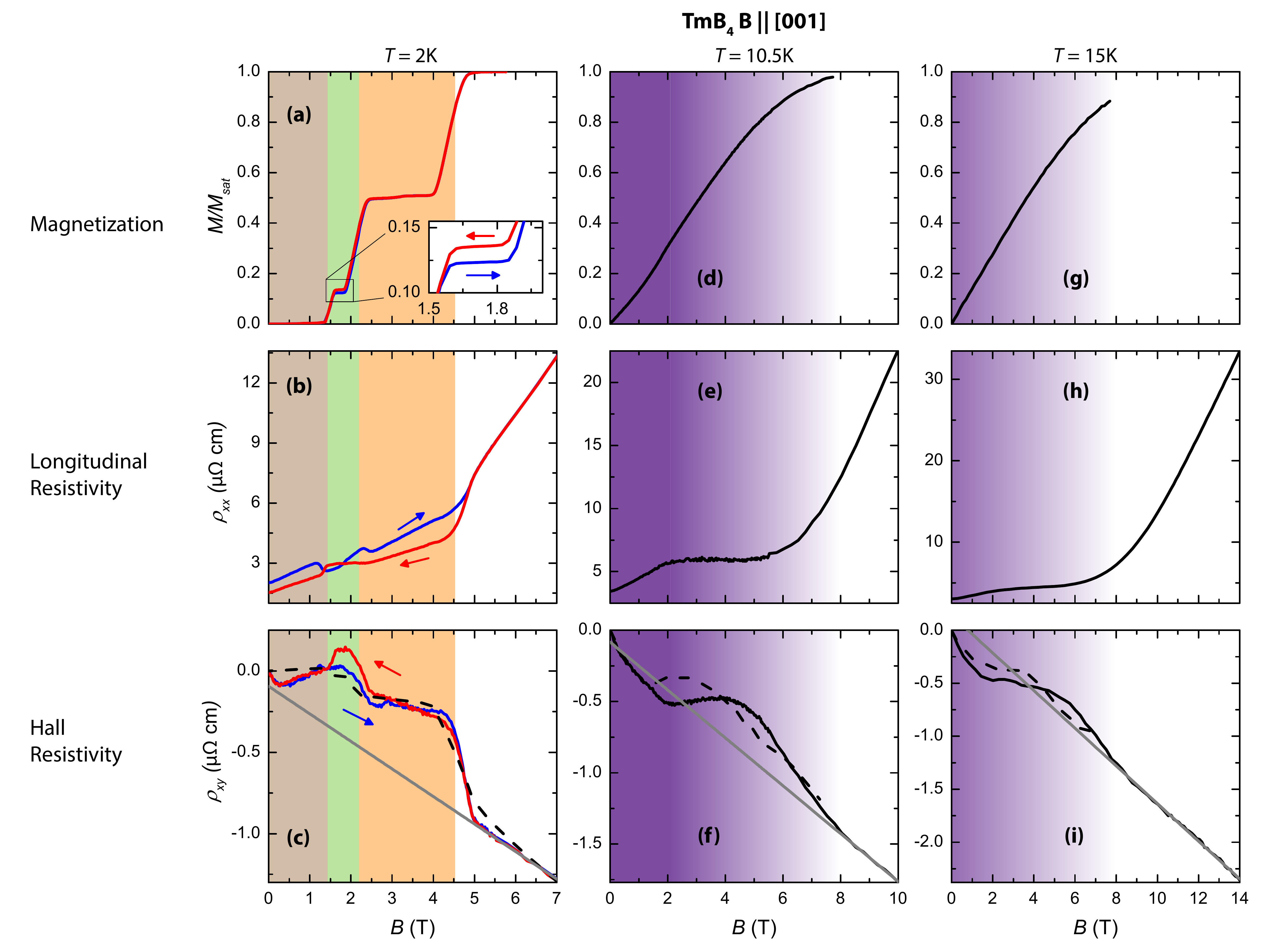}
	\caption{{\textbf{(a)-(c)} $M$, $\rho_{xx}$ and $\rho_{xy}$ at 2K. \textbf{(d)-(f)} $M$, $\rho_{xx}$ and $\rho_{xy}$ at 10.5K. \textbf{(g)-(i)} $M$, $\rho_{xx}$, and $\rho_{xy}$ at 15K. The dashed lines are best fits to conventional AHE theories (Eqn. \ref{Eqn_AHE}) and the solid grey lines {are linear fits to $\rho_{xy}$ above saturation.} At 2K, the best fit is to the downsweep. The colored backgrounds correspond to different magnetic phases (Fig. \ref{Fig_phase_diag}(c)).}}
\label{Fig_Hall}
\end{figure*}

\textit{Results} - An examination of the in plane and out of plane longitudinal resistivities ($\rho_{xx}$ and $\rho_{zz}$, Fig \ref{Fig_temp}(a)) reveals two key features. First, $\rho_{xx}$ and $\rho_{zz}$ show a significant drop at $T_{N1}$ and $T_{N2}$ due to decrease in scattering from disordered spins. Second, both $\rho_{xx}$ and $\rho_{zz}$ are very similar in magnitude and $T$-dependence. The second result is in sharp contrast with the assumption of \ce{TmB4} being a quasi 2D system \cite{Chang2009, Suzuki2009, Suzuki2010, Dublenych2012}. To rule out a possible misalignment, we confirmed the orientation of the crystal used for $c$-axis transport after the experiments \cite{Suppl}. We conclude that \ce{TmB4} is an electronically 3D system. This result is expected from the 3D crystal structure: the smallest distance between the Tm ions along the c-axis is 3.987{\AA} while the corresponding in plane distance is 3.64{\AA} \cite{Fisk1972}. Further support comes from band structure calculations \cite{Yin2008} {and quantum oscillation measurements on the related compound \ce{YB4} \cite{Tanaka1985}}, which show that the Fermi surface is 3D.

The isotropic nature of the resistivity implies that the out of plane magnetic interactions between Tm spins are non-negligible in comparison to the in plane interactions $\mathcal{J}_1$ and $\mathcal{J}_2$. Future theoretical models must take this result into consideration. We suggest that an anisotropic Kondo lattice model, similar to that used for $\beta$-\ce{YbAlB4} \cite{Nevidomskyy2009}, may be more appropriate for \ce{TmB4}, although further experiments are needed to establish such a picture.

The in plane Hall resistivity ($\rho_{xy}$, Fig \ref{Fig_temp}(b)) decreases at high temperature but shows a sharp upturn at $T_{N1}$ and a change of slope at $T_{N2}$. {To investigate this unusual behavior in $\rho_{xy}$}, we measured the magnetic field dependence of $M$, $\rho_{xx}$ and $\rho_{xy}$ at three temperature regimes: $T < T_{N2}$ (2K), $T_{N2} < T < T_{N1}$ (10.5K) and $T > T_{N1}$ (15K), shown in Fig.~\ref{Fig_Hall}.

The magnetization at 2K, shown in Fig.~\ref{Fig_Hall}(a) as a function of magnetic flux density $B = \mu_0 H+M$ \cite{Suppl}, displays the previously reported plateau structure \cite{Iga2007, Gabani2008, Siemensmeyer2008}. $\rho_{xx}$ at 2K (Fig. \ref{Fig_Hall}(b)), shows features at the magnetic transitions indicating a strong influence of the magnetic structure on $\rho_{xx}$. Similar features have been observed in other metallic magnets such as \ce{SrCo6O11} \cite{Ishiwata2007} and \ce{RNi2Ge2} \cite{Budko1999}. Suprisingly, $\rho_{xx}$ shows a strong hysteresis at all magnetic fields below saturation, including zero field, even though the magnetization shows a noticeable hysteresis only at the fractional plateau.

The vanishing of hysteresis in MR above saturation allows us to exclude nonmagnetic explanations such as structural defects and extrinsic impurities.  Hysteretic MR has previously been observed in phase separated perovskite manganites \cite{Krivoruchko2008} and ferromagnets such as \ce{Fe_{1/4}TaS2} \cite{Checkelsky2008}, where it is the result of a change in the bulk magnetic structure. The presence of a hysteresis in MR with no corresponding hysteresis in magnetization is counterintuitive (because the lack of hysteresis in the magnetization {suggests} that the magnetic structure remains the same). We return to this result later.

We now examine the Hall resistivity in \ce{TmB4}. Conventionally, the Hall resistivity of a magnetic material can be decomposed into its ordinary contribution which depends on {$B$} \cite{Suppl}, and an anomalous contribution {which depends on $M$ and the scattering rate (through $\rho_{xx}$)} \cite{Nagaosa2010}:
\begin{eqnarray}
\rho_{xy} = R_0 {B} + (a\rho_{xx} + b\rho_{xx}^2)M,
\label{Eqn_AHE}
\end{eqnarray}
where $R_0$ is the ordinary Hall coefficient and $a$ and $b$ are constants. The second term ($\rho_{xy} \sim \rho_{xx} M$) is due to the skew scattering mechanism \cite{Smit1955, Smit1958}, while the third term ($\rho_{xy} \sim \rho_{xx}^2 M$) is a combination of intrinsic AHE and side jump mechanisms \cite{Jungwirth2002, Onoda2002, Berger1970}. By comparing our data to Eqn.~\ref{Eqn_AHE}, we can test if the AHE in \ce{TmB4} can be explained by conventional theories. While some of the magnetic phases, especially the fractional plateau phase, extend over a narrow $H$-range to allow a definite comparison, our conclusions remain unaffected.

$\rho_{xy}$ at 2K (Fig. \ref{Fig_Hall}(c)) consists of regions of linear behavior separated by sharp jumps and shows hysteresis between 1.4T and 2.5T. We notice that $\rho_{xy}$ does not scale with magnetization. As we go from the N\'{e}el phase (brown) to the fractional plateau phase (green), the magnetization increases and $\rho_{xy}$ shows a corresponding increase. However, as we reach the half plateau (orange), $\rho_{xy}$ drops. Saturation (white) leads to an even larger drop in $\rho_{xy}$. {Moreover, $\rho_{xy}$ is linear above saturation despite the presence of a large, nonsaturating MR. {This result shows that ordinary contributions to $\rho_{xy}$ dominate above saturation and conventional contributions to AHE are negligibly small ($a \simeq 0$, $b\simeq 0$ in Eqn \ref{Eqn_AHE}).} A best fit of the down sweep to Eqn 1, while showing good agreement between 2T and 4T, deviates significantly below 2T and is strongly nonlinear above saturation (Fig. S7 in \cite{Suppl}).

The magnetic and transport properties of MP1 are qualitatively similar to those of MP2 \cite{Suppl} and we focus our analysis on the latter. At 10.5K, the long-range modulation of MP2 disappears at 1.6T and the magnetization saturates at $\sim 7$T (Fig. \ref{Fig_Hall}(d)). $\rho_{xy}$ shows a sharp kink at 1.6T, then a broad hump at $\sim 4$T before finally becoming linear above saturation  (Fig. \ref{Fig_Hall}(f)). {Considering the behavior of $M$ and $\rho_{xx}$  (Fig. \ref{Fig_Hall}(e)), both of which do not show a hump, conventional contributions to AHE cannot lead to the observed $\rho_{xy}$.} Despite the presence of a strong MR above saturation, $\rho_{xy}$  is linear, indicating that conventional contributions to AHE can be neglected. A best fit of $\rho_{xy}$ to Eqn \ref{Eqn_AHE} deviates strongly from the measured data (Fig. \ref{Fig_Hall}(f)).

At $T > T_{N1}$, no long-range magnetic order is present and $M$ (Fig. \ref{Fig_Hall}(g)) {increases smoothly until the maximum measured field}. Both $\rho_{xx}$ and $\rho_{xy}$ at 15K (Figs. \ref{Fig_Hall}(h-i)) are very similar to the corresponding curves at 10.5K, despite the absence of long range order at 15K. $\rho_{xy}$ shows a kink at 1T and a broad hump at $\sim 5$T before becoming linear above saturation. Using the same arguments as those at 10.5K, {we conclude that conventional contributions to AHE are negligibly small at 15K} and a best fit of $\rho_{xy}$ to Eqn. \ref{Eqn_AHE} deviates strongly from the measured data (Fig. \ref{Fig_Hall}(i)).

{An unusual feature common to the $\rho_{xy}$ data at all three temperatures is the non-zero y-intercept of the linear fit above saturation. However, the slope of linear fit to the $\rho_{xy}$ data is comparable at all three temperatures (Sec. IX in \cite{Suppl}). The carrier concentration calculated at 2K matches well with the value at 300K (where no AHE is expected to be present) as well as the experimentally measured value on the non-magnetic compound \ce{YB4} (Sec. IX in \cite{Suppl}). This correspondence suggests that the high-field behavior of $\rho_{xy}$ is the sum of a linear contribution from ordinary Hall effect and a constant term.}

\textit{Discussion} - The MR of \ce{TmB4} shows strong hysteresis at 2K despite the absence of corresponding hysteresis in the magnetization. We suggest that subtle changes occur in the magnetic structure of \ce{TmB4} that strongly influence the MR but not the bulk magnetization. Neutron scattering experiments have shown that the magnetic structure in the modulated and the plateau phases consists of stripes or domains \cite{Siemensmeyer2008, Gabani2008, Michimura2009}. However, the microscopic structure at the domain walls is unknown. The domain walls could contain unusual magnetic structures or disordered spins or both, {a possibility not considered in previous studies on \ce{TmB4}}. Changes in those structures can lead to a hysteretic MR while leaving the bulk magnetization unaffected.

By considering the behavior of Hall resistivity above saturation, we find that conventional contributions to AHE are negligibly small in \ce{TmB4}. {Therefore, all deviations from the ordinary, linear field dependence are due to unconventional mechanisms.} One possibity is topological Hall effect (THE) where conduction electrons moving through a noncoplanar structure accumulate a Berry phase due to net spin chirality leading to a Hall contribution. However, neutron scattering experiments on \ce{TmB4} have not found any evidence for a global noncoplanar structure \cite{Siemensmeyer2008, Michimura2009}. {We suggest} that noncoplanar structures {could} arise at domain walls which in turn lead to both hysteretic MR and THE.  {Further experiments are necessary to confirm this hypothesis.} {Above saturation, the magnetic structure is coplanar and any potential THE contributions must be zero.} In contrast, our data shows that a constant term is present. Therefore, additional contributions to AHE must be present. Other possibilities are AHE arising from phonons and spin waves \cite{Nagaosa2010, Oda2001}. Further work is necessary to determine if they can account for the measured $\rho_{xy}$ in \ce{TmB4}.

In conclusion, we discovered that \ce{TmB4}, and likely other \ce{RB4}, are electronically 3D systems and future {theoretical} models must take this {result} into consideration. {Our hysteretic MR results suggest that complex structures arise at magnetic domain walls that strongly affect the transport properties. Our Hall resistivity results show the presence of AHE. Further analysis reveals that conventional contributions to the AHE are negligible and hence unconventional contributions must be present.} A combination of high resolution neutron scattering, microscopic experiments and theoretical modelling are required to {determine the magnetic structure and the origin of unconventional AHE in \ce{TmB4}}.

\begin{acknowledgments}
We thank Y. Ozaki for technical assistance in crystal alignment. SSS thanks Tanmoy Das and Anjan Soumyanarayanan for helpful discussions. Work in Singapore was supported by grant MOE2011-T2-1-108 from the Ministry of Education, Singapore and the National Research Foundation (NRF), NRF-Investigatorship (NRF-NRFI2015-04). Work at Ames Laboratory was supported by the U.S. Department of Energy, Office of Basic Energy Science, Division of Materials Sciences and Engineering. Ames Laboratory is operated for the U.S. Department of Energy by Iowa State University under Contract No. DE-AC02-07CH11358. Work at  UCSC was supported by the U.S. Department of Energy, Office of Science, Basic Energy Sciences under Award \#FG02-06ER46319.
\end{acknowledgments}

\bibliographystyle{apsrev}
\bibliography{TmB4-ref}

\end{document}


\title{Hysteretic magnetoresistance and unconventional anomalous Hall effect in the frustrated magnet \ce{TmB4}}
\author{Sai~Swaroop~Sunku,$^1$  Tai~Kong,$^2$ Toshimitsu~Ito,$^3$ Paul~C.~Canfield,$^2$ B.~Sriram~Shastry,$^4$ Pinaki~Sengupta,$^1$ and Christos~Panagopoulos$^1$}
\affiliation{$^1$Division of Physics and Applied Physics, School of Physical and Mathematical Sciences, Nanyang Technological University, 21 Nanyang Link, Singapore 637371}
\affiliation{$^2$Ames Laboratory, U.S. DOE and Department of Physics and Astronomy, Iowa State University, Ames, Iowa 50011, USA}
\affiliation{$^3$National Institute of Advanced Industrial Science and Technology (AIST), Tsukuba, Ibaraki 305-8562, Japan}
\affiliation{$^4$Physics Department, University of California, Santa Cruz, CA 95064, USA}

\maketitle

\begin{center}
{\large \textbf{SUPPLEMENTAL MATERIAL}}
\end{center}

\setcounter{equation}{0}
\setcounter{figure}{0}
\setcounter{table}{0}
\setcounter{page}{1}
\makeatletter
\renewcommand{\theequation}{S\arabic{equation}}
\renewcommand{\thefigure}{S\arabic{figure}}
\renewcommand{\bibnumfmt}[1]{[S#1]}
\renewcommand{\citenumfont}[1]{S#1}
\renewcommand{\thetable}{S\arabic{table}}

\section{Detailed Methods}

\subsection{Crystal growth and alignment}
Our experiments were performed on TmB$_4$ single crystals synthesised by the solution growth method using an Al flux. Bulk starting elements with a ratio of Tm:B:Al = 0.125 : 0.75 : 50 were put into an alumina crucible, which was heated up to 1475$^{\circ}$C and cooled down to 750$^{\circ}$C over a period of ten days in a continuous flow of high-purity argon and then quenched to room temperature via furnace cooling. The growth was then taken out from the furnace at room-temperature and re-sealed into a silica ampoule. Single crystals of \ce{TmB_4} were separated from the remaining liquid in a centrifuge after heating the ampoule back up to 750$^{\circ}$C. The crystals were oriented using X-ray diffraction in the Laue geometry with an error of less than $\pm 5^{\circ}$. For transport experiments, the crystals were cut with a tungsten wire saw into cuboids with faces along [100] or [001].\\

A total of four crystals (Samples 1-4) were used in our experiments. Samples 1 and 2 were used for magnetization experiments. All reported magnetization data was obtained from Sample 2 and key features were verified with Sample 1. Samples 3 and 4 were cut and used for in-plane and c-axis transport experiments, respectively.

\subsection{Experimental techniques}
Magnetic field was always applied along [001]. DC magnetization measurements were performed in a Quantum Design MPMS XL SQUID magnetometer. The magnetic field for SQUID measurements was limited to 7T. \\

For transport experiments, electrical contacts to the sample were made by attaching 25$\mu$m/50$\mu$m gold wires with silver epoxy (DuPont 6838 or EpoTek E4110), which ensured ohmic contacts to the sample. Resistivity and Hall measurements were performed using a commercial measurement system (Quantum Design PPMS) using standard 4-probe AC transport techniques. For in-plane measurements, the current was applied along [100]. To correct for contact misalignment, measurements were performed at both positive and negative fields and the data was symmetrized accordingly (Section V). For out-of-plane resistivity measurements, the current was applied along [001]. {All measurements were performed with a current 1.8mA and 5mA. No difference was found in the value of Hall resistivity.}\\

The small size of the samples results in considerable uncertainties in determing the distance between the electrical contacts. As a result, an error bar of 20\% is present on the absolute values of all transport quantities. However, this error does not affect any of our conclusions.

\pagebreak

\subsection{Field protocol}
The magnitude of the magnetization at the fractional plateau is known to vary with the field history \cite{Siemensmeyer2008}. A protocol was developed that reproduced the same magnetization curve at 2K when the measurement is repeated:

\begin{enumerate}
\item Cool down to 2K in zero magnetic field from above $T_{N1}$ (Zero-Field-Cool)
\item Sweep the magnetic field up to 5T to reach the saturation phase and sweep down to 0T (Pole)
\item Perform measurements, making sure to sweep the field up to at least 5T at the end of the sweep. (Measure)
\end{enumerate}

The second step (poling) was found to be necessary to obtain reproducible curves. All further measuerments after poling were reproducible, provided the field was always swept up to at least 5T (step 3). This protocol was used for both magnetization and transport measurements performed at 2K. No hysteresis was observed at 10.5K in both magnetic and transport properties.

\pagebreak

{\section{Demagnetization corrections}}

{
In a magnetic material, an external magnetic field $H_{app}$ induces a demagnetization field $H_{demag}$ which is proportional to the material's magnetization: $H_{demag} = NM$ and $N$ is the demagnetization factor. The effective field in the material, $H_{eff}$ is given by}

{
\begin{equation}
H_{eff} = H_{app} - N M.
\end{equation}
}

{
Our magnetization measurements were performed on Sample 2, which is an irregularly shaped, as-grown crystal. We approximated it as a sphere of diameter 1.2mm and its demagnetization factor is $N_{S2} = 1/3$. Our ab-plane transport experiments were performed on Sample 3 which is a rectangular prism of dimensions 0.516mm x 0.434mm x 0.226mm. We used the formula of Aharoni \cite{Aharoni1998} to calculate its demagnetization factor and obtained $N_{S3} = 0.506$.\\}

\begin{figure}[h!]
	\includegraphics[trim=0cm 0cm 0cm 0cm, width=0.65\linewidth]{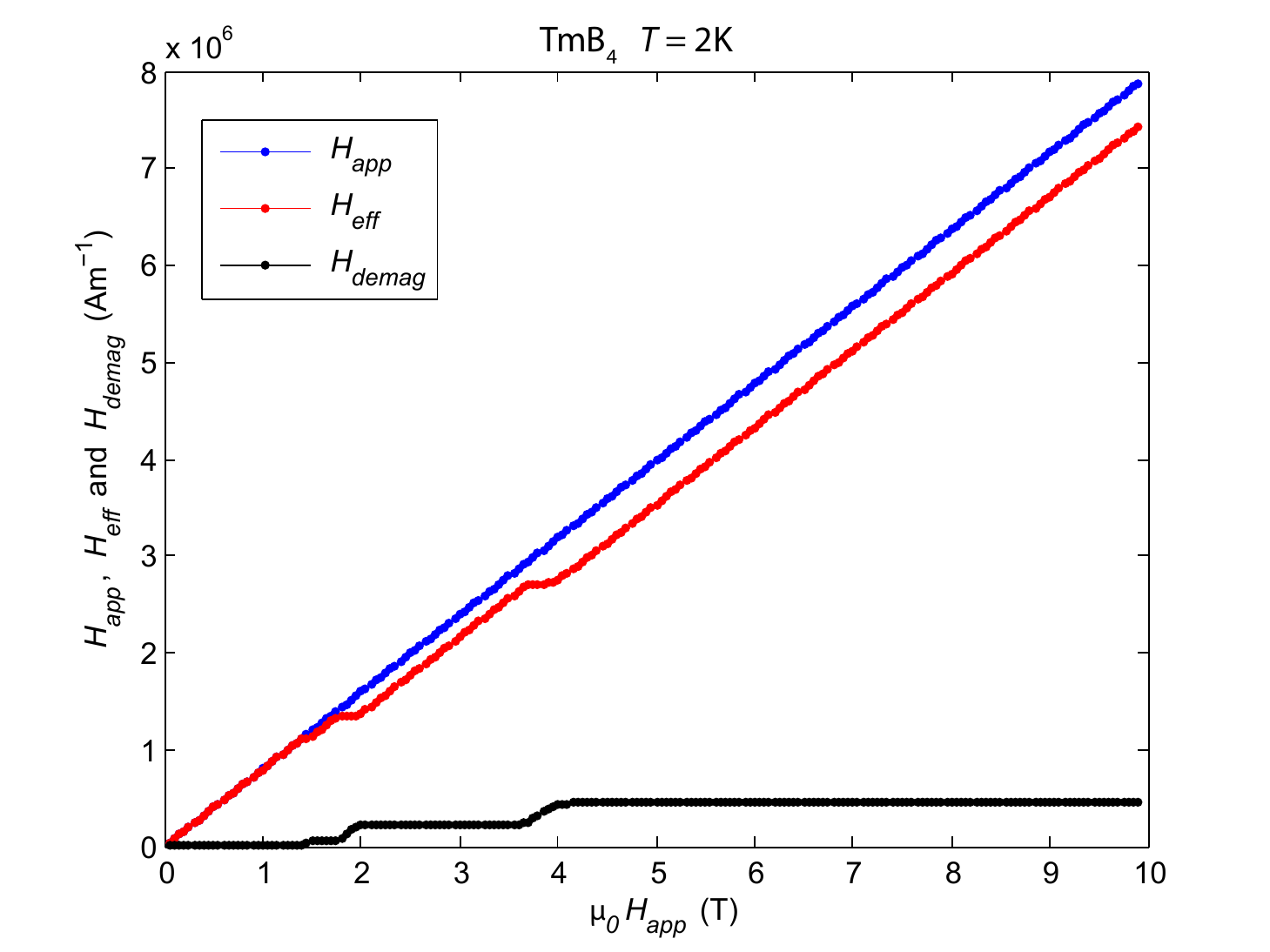}
	\caption{$H_{app}$, $H_{eff}$ and $H_{demag}$ for a downsweep at 2K.}
\label{Fig_demag_field}
\end{figure}

{
Figure \ref{Fig_demag_field} shows $H_{app}$, $H_{eff}$ and $H_{demag}$ for a downsweep at $T = 2$K. $H_{demag}$ follows the profile of the magnetization and its magnitude is about 6\% of $H_{app}$ when $\mu_0 H_{app} = 10$T.}

\begin{figure}[h!]
	\includegraphics[trim=5cm 7.5cm 5cm 7.5cm, width=0.6\linewidth]{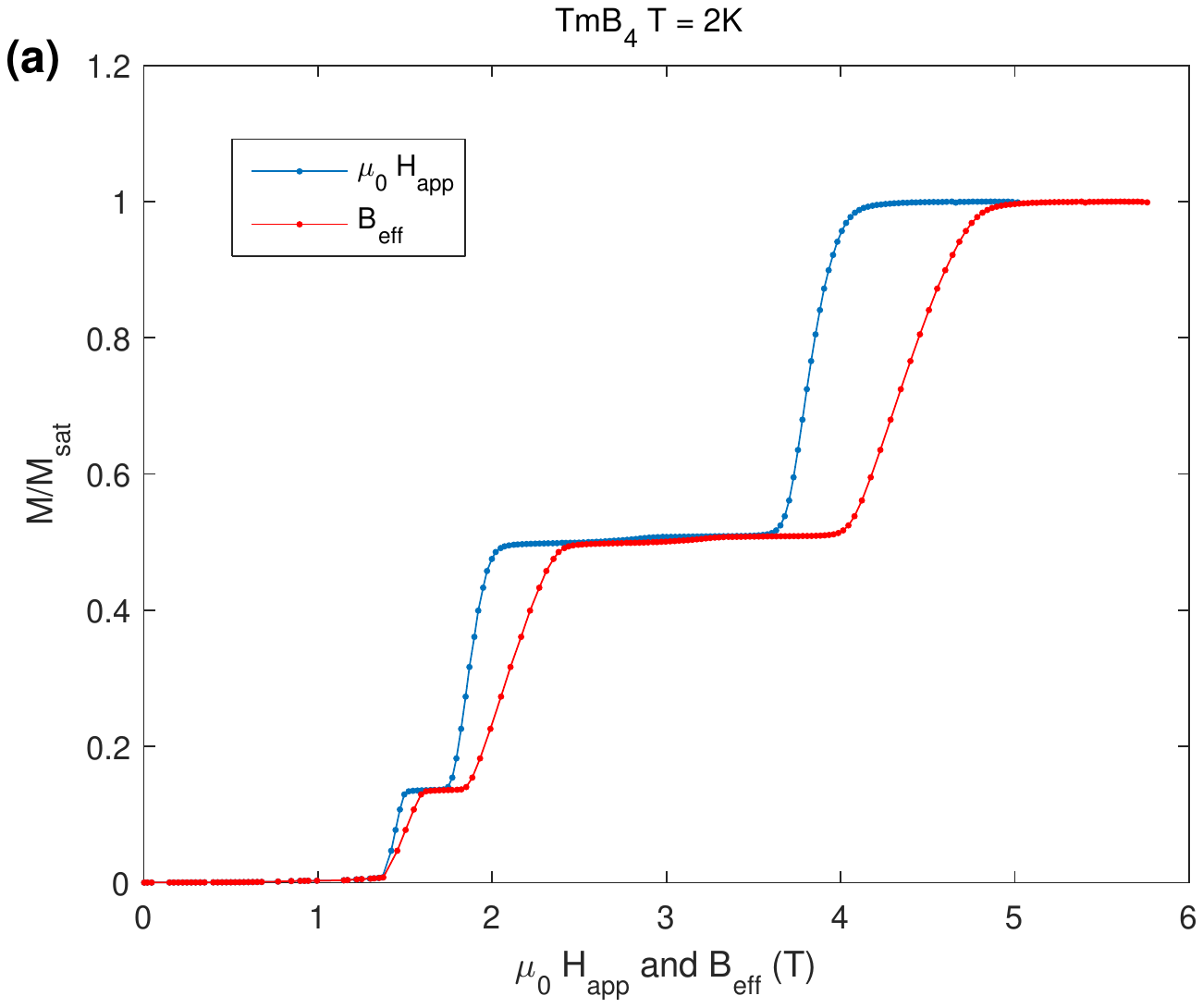}\vspace{20pt}
	\includegraphics[trim=5cm 7.5cm 5cm 9cm, width=0.6\linewidth]{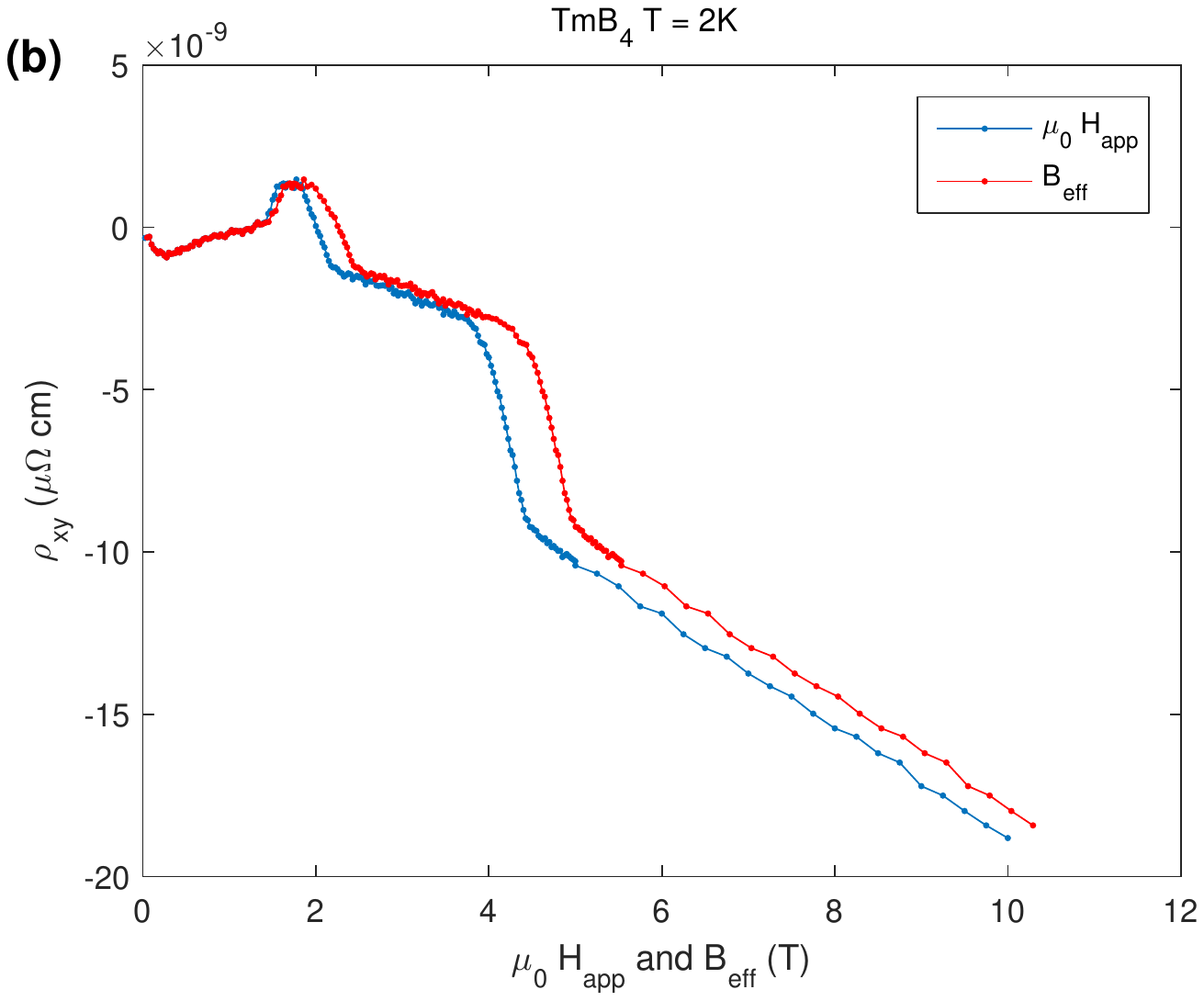}
	\caption{\textbf{(a)} Magnetization data during a downsweep at 2K plotted against $H_{app}$ and $H_{eff}$. \textbf{(b)} Hall resistivity during a downsweep at 2K plotted against $H_{app}$ and $H_{eff}$.}
\label{Fig_demag_data}
\end{figure}

{
The magnetic flux density $B$ is related to the magnetic field as $B = \mu_0(H+M)$. The effective flux density $B_{eff}$ is given by}
\begin{equation}
{B_{eff} = \mu_0(H_{eff}+M) = \mu_0(H_{app} + (1-N)M).\\}
\end{equation}

To correct for the demagnetization, we calculated $B_{eff}$ from $H_{app}$ and $M$ for each measurement. Figure \ref{Fig_demag_data}(a) shows $M$ as a function of $\mu_0 H_{app}$ and $B_{eff}$ for a downsweep at 2K. To correct the transport data, we interpolated the magnetization data and determined the magnetization at every field point of the transport field sweeps. We then used the demagnetization factor of the transport sample to obtain $B_{eff}$. Figure \ref{Fig_demag_data}(b) shows $\rho_{xy}$ as a function of $\mu_0 H_{app}$ and $B_{eff}$ for a downsweep at 2K.
{In the rest of the manuscript, we refer to $H_{eff}$ as $H$ and $B_{eff}$ as $B$ for simplicity, except when stated otherwise.}\\

\clearpage

\section{Phase diagram}

\begin{figure}[h!]
	\includegraphics[trim=1cm 0cm 1cm 1cm, width=0.45\linewidth]{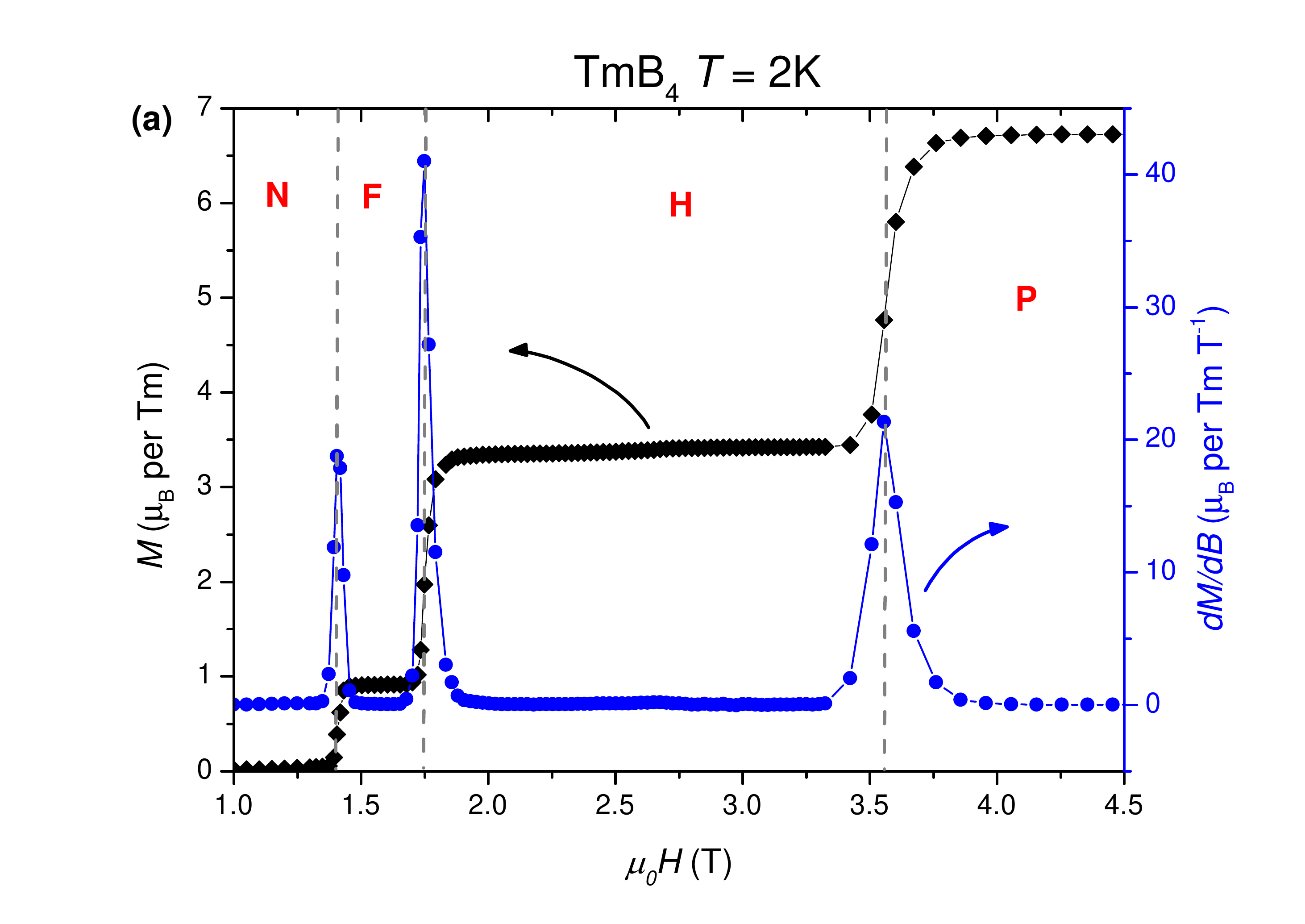}
	\includegraphics[trim=1cm 0cm 1cm 1cm, width=0.45\linewidth]{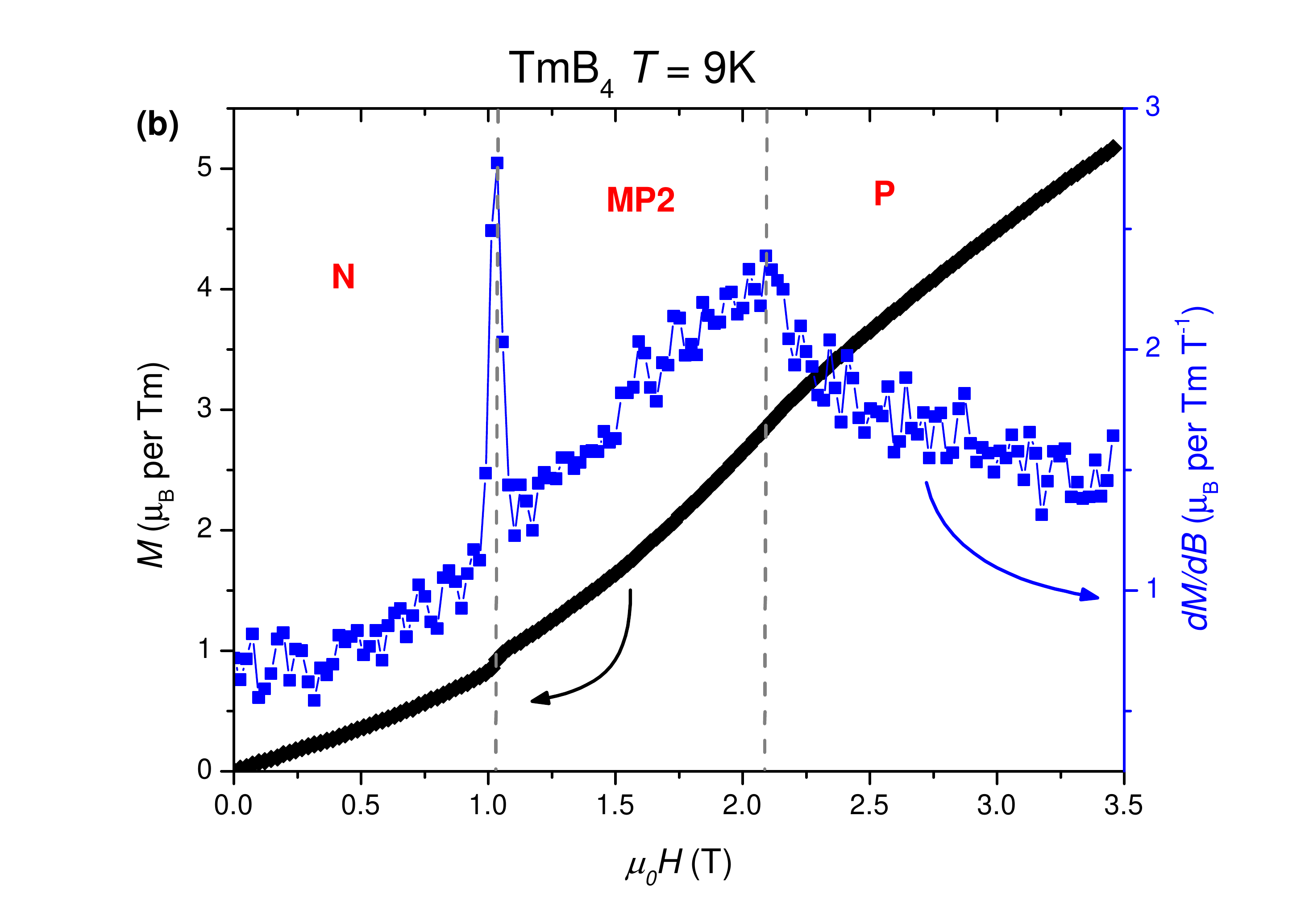}
	\includegraphics[trim=1cm 1cm 1cm 0cm,width=0.45\linewidth]{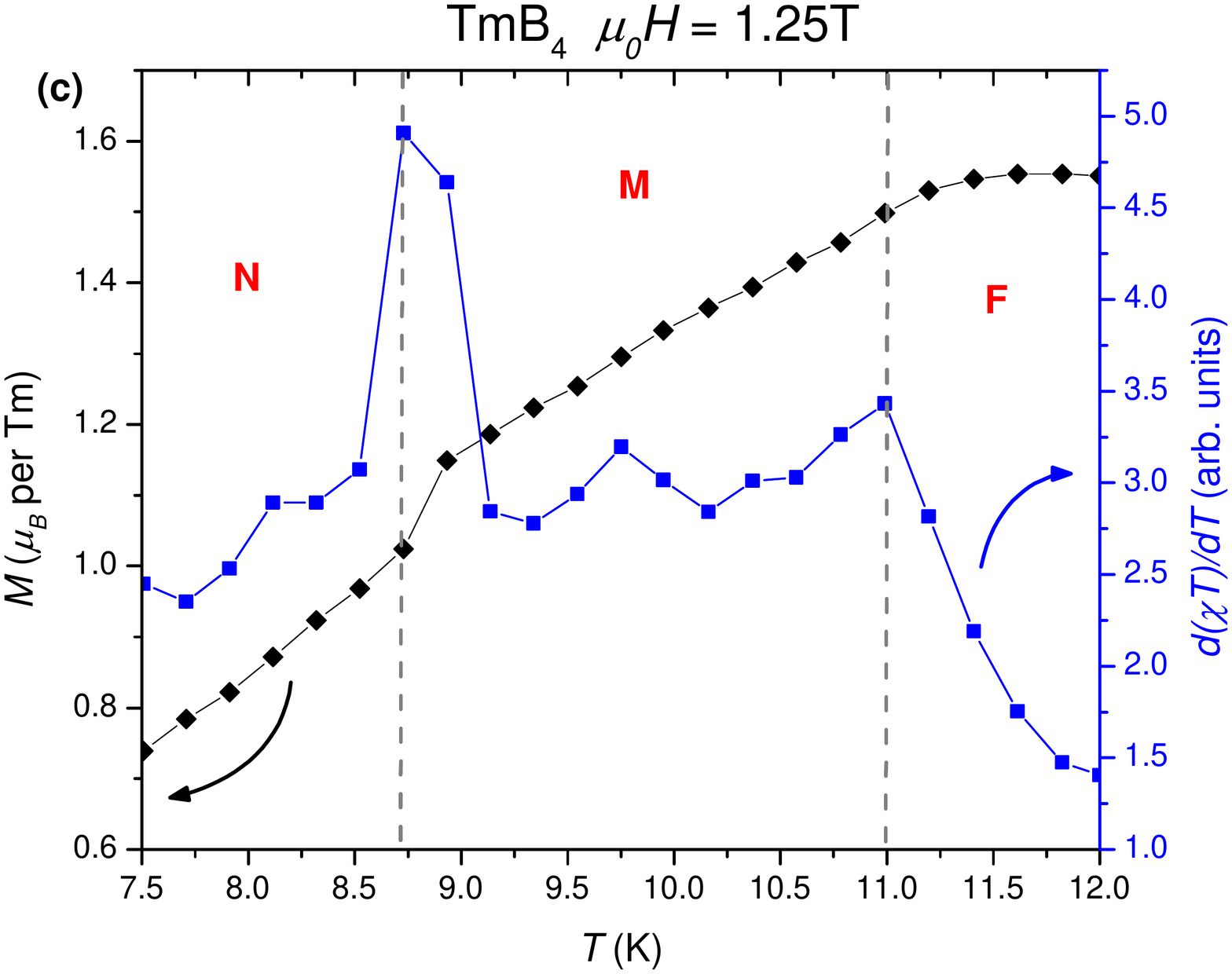}
	\includegraphics[trim=1cm 1cm 1cm 0cm,width=0.45\linewidth]{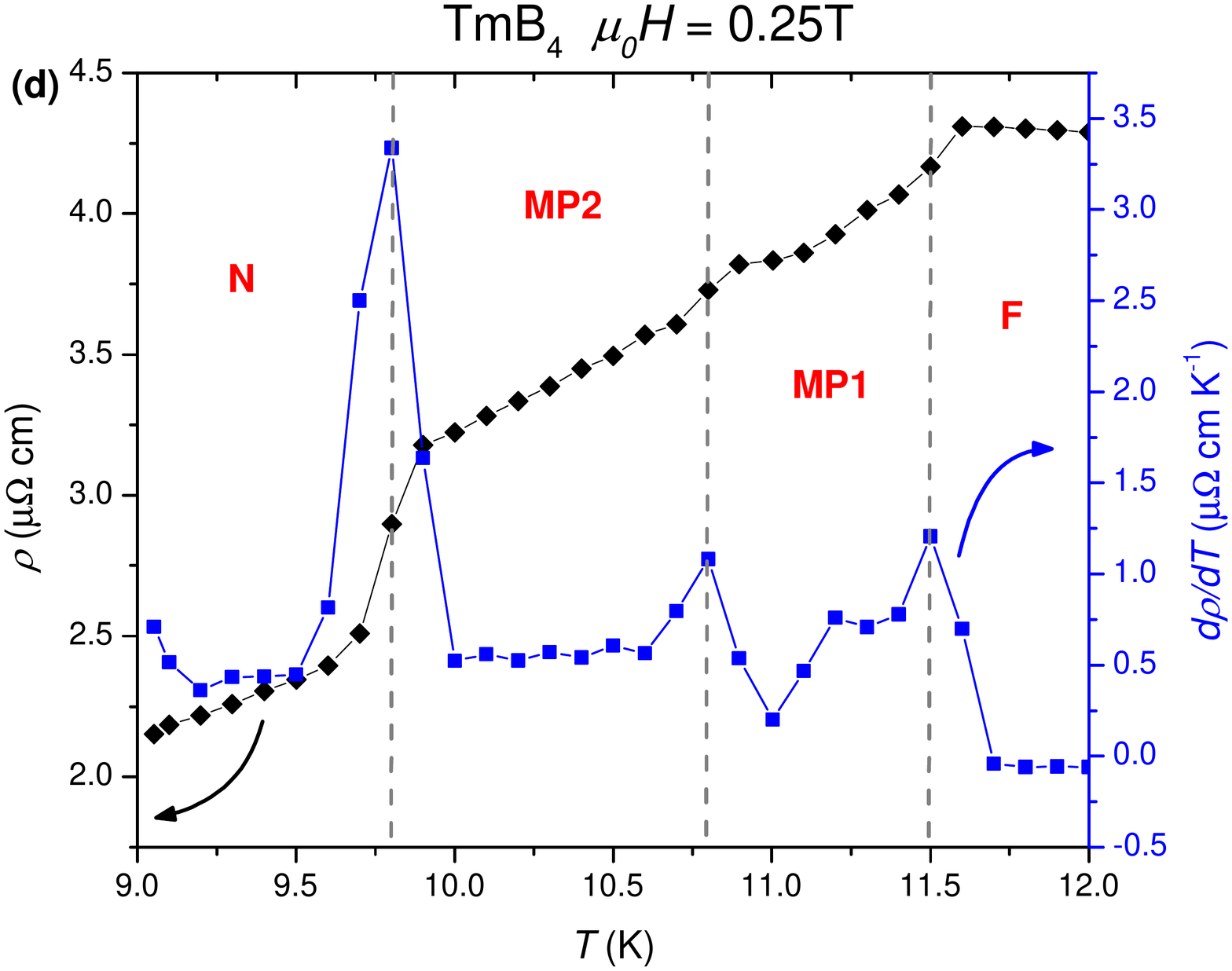}
	\caption{Representative raw data of $dM/dB$, $dM/dT$ and $d\rho_{xx}/dT$ calculated using a three-point derivative algorithm. The phase transitions appear as peaks, which are used to obtain the phase diagram in Figure 1 of the main text. N is the N\'{e}el phase, F is the fractional plateau, H is the half plateau, P is the paramagnetic phase, M is Modulated Phase, MP1 and MP2 are Modulated Phase I and Modulated Phase II respectively, and F is the fluctuating phase. We estimate an error bar of 10\% on the absolute value of magnetization by comparing the saturation magnetization of different samples and measurement runs. The source of the error is probably a slight misalignment of the crystal during measurement.  \textbf{(a)} $M$ and $dM/dB$ at $T = 2$K. The saturation magnetization is slightly smaller than the expected value of 7$\mu_B$/Tm but is within the estimated error bar. \textbf{(b)} $M$ and $dM/dB$ at $T = 9$K. \textbf{(c)} $M$ and $d(\chi T)/dT$ at $B = 1.25$T. The transition between MP1 and MP2 is not visible in the magnetization curves. \textbf{(d)} $\rho_{xx}$ and $d\rho_{xx}/dT$ at $B = 0.25$T.}
\label{Fig_phase_diag_raw}
\end{figure}

We obtained the points on the phase diagram by locating the peaks in the following derivatives: $dM/dB$, $d(\chi T)/dT$ and $d\rho_{xx}/dT$. We use $\mu_0 H$ instead of $B$ to allow comparison with previous work \cite{Siemensmeyer2008, Michimura2009, Wierschem2015}. Representative curves are shown in Figure \ref{Fig_phase_diag_raw}. \\

\pagebreak

\section{Orientation of the crystal used for c-axis transport}

\begin{figure}[h!]
	\includegraphics[trim=2cm 1cm 1cm 1cm, width=0.5\linewidth]{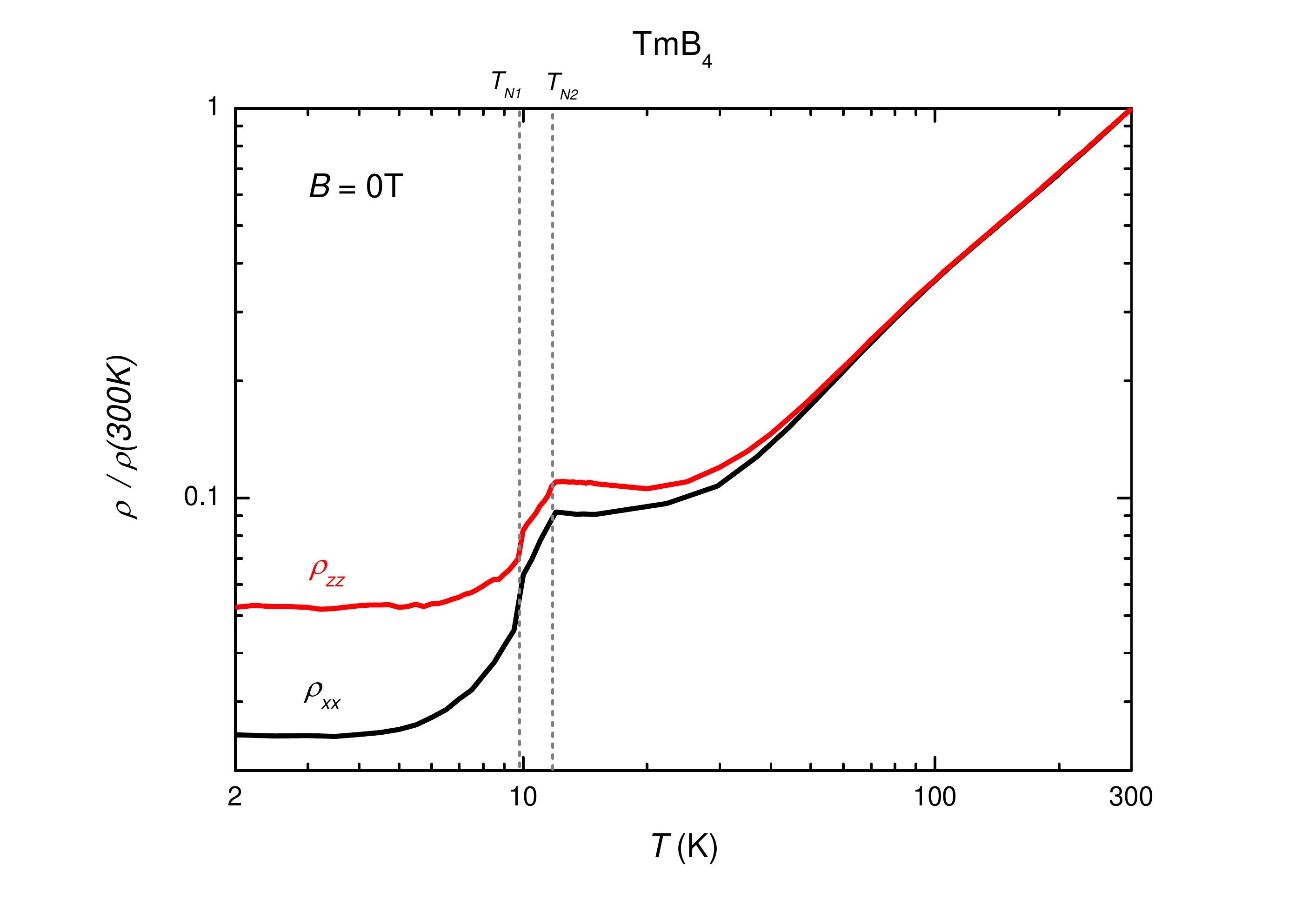}
	\caption{Plot of $\rho_{xx}/\rho_{xx}$(300K) and $\rho_{zz}/\rho_{zz}$(300K) of \ce{TmB4} showing the very similar temperature dependence.}
\label{Fig_rho_300K}
\end{figure}

In Figure 2(a) of main text, we show $\rho_{xx}$ and $\rho_{zz}$ of \ce{TmB4} and find that they are very similar in magnitude and $T$-dependence. Figure \ref{Fig_rho_300K} confirms this result. One possible experimental error that could lead to this behaviour is a mistake in the orientation of the crystal such that the measured $\rho_{zz}$ is actually $\rho_{xx}$. We checked for this scenario by taking a Laue photograph of the crystal after the transport measurements were completed.\\

\begin{figure}[h!]
	\includegraphics[trim=0cm 3cm 0cm 2cm, width=0.95\linewidth]{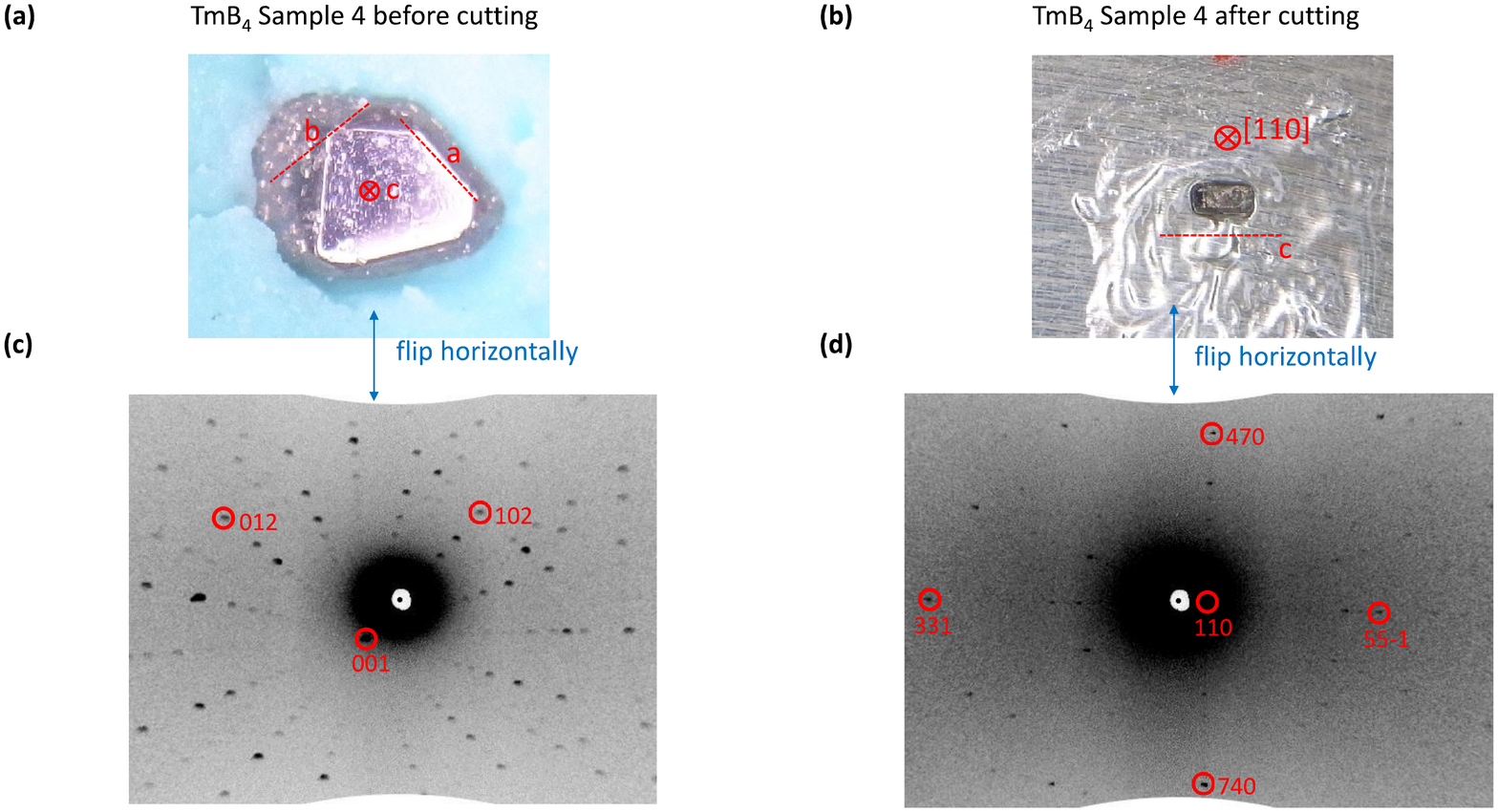}
	\caption{\textbf{(a)} and \textbf{(b)} show pictures of Sample 4 before and after cutting. \textbf{(c)} and \textbf{(d)} show the corresponding Laue photographs.}
\label{Fig_c-laue}
\end{figure}

Figure \ref{Fig_c-laue}(a) shows a picture of pristine Sample 4 crystal. Figure \ref{Fig_c-laue}(b) shows a picture of the piece that was used for transport experiments. The electrical current was applied along the long axis. Figures \ref{Fig_c-laue}(c) and \ref{Fig_c-laue}(d) show the Laue photographs of the pristine and cut crystal respectively. The Laue photographs confirm that the long axis of the cut crystal is the c-axis.

\pagebreak

{
\section{Raw transverse voltage}}

\begin{figure}[h!]
	\includegraphics[trim=1cm 1cm 2cm 2cm, width=0.8\linewidth]{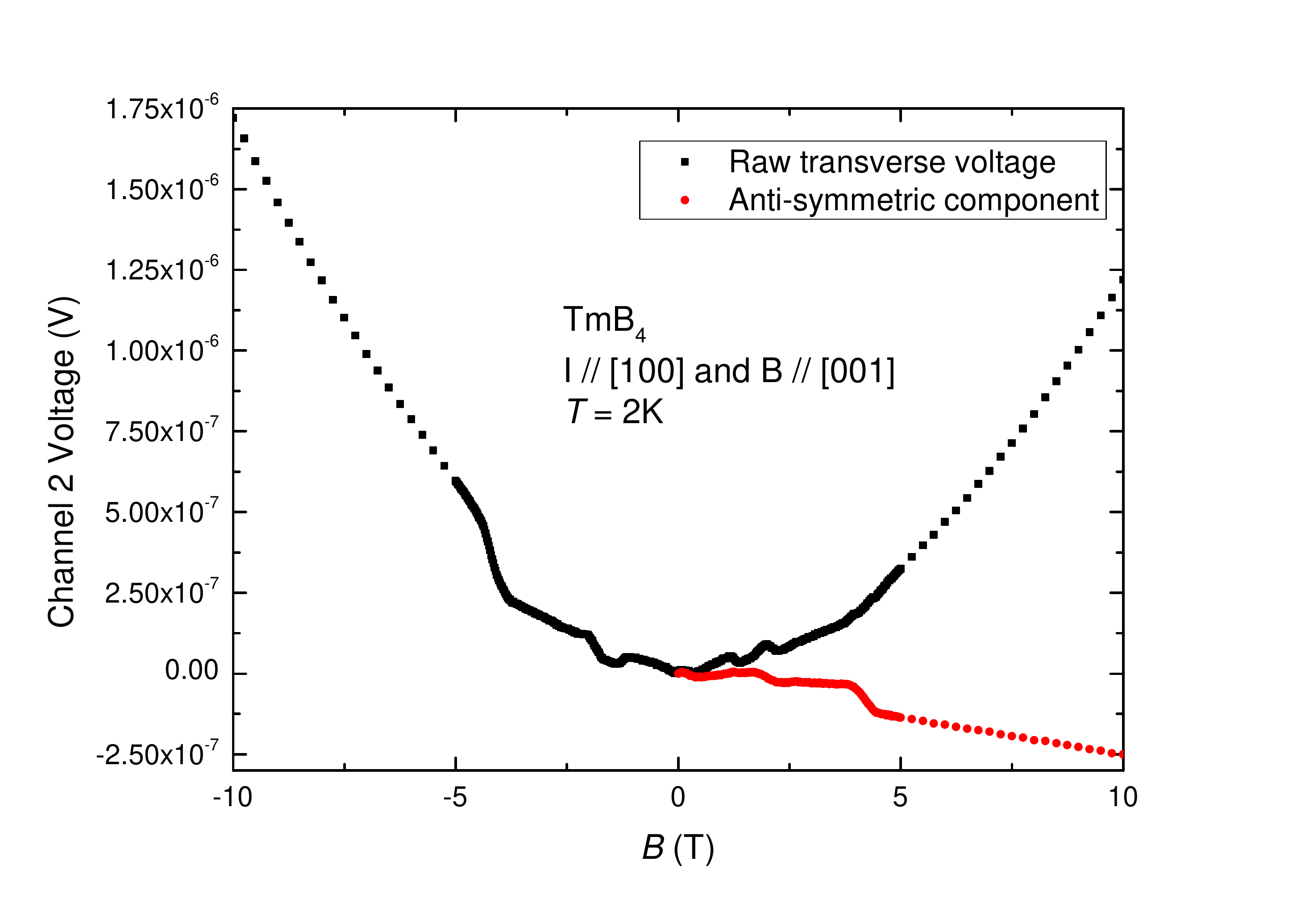}
	\caption{Raw transverse voltage and its anti-symmetric component for a typical field sweep at $T =$ 2K. The above curve has not been corrected for demagnetization.}
\label{Fig_raw_trans}
\end{figure}

{
Figure \ref{Fig_raw_trans} shows the raw transverse voltage and its anti-symmetric component for a typical field sweep at 2K. The anti-symmetric component is typically 20\% of the total signal or greater. The magnitude of the anti-symmetric component is in the 100nV regime (reaching a maximum of 250nV at 10T) which is well above our typical noise level of $\sim$5nV. Therefore, the magnetoresistance contamination in the Hall resistivity measurements does not affect the conclusions presented in the manuscript.\\}

\pagebreak
{\section{Conventional AHE fit at high magnetic fields}}

\begin{figure}[h!]
	\includegraphics[trim=1cm 1cm 1cm 1cm, width=0.75\linewidth]{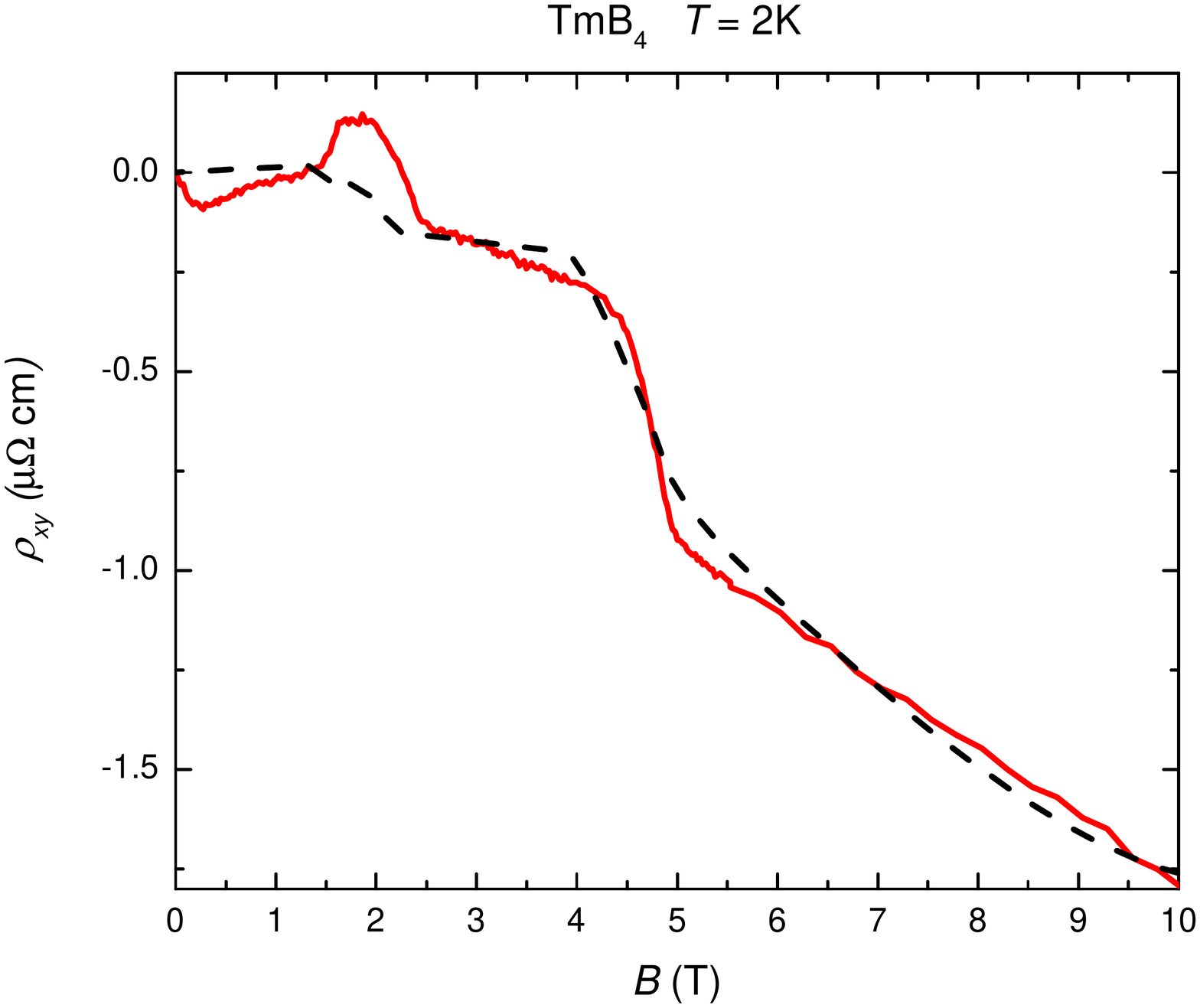}
	\caption{High-field Hall resistivity in \ce{TmB4} at 2K. The dashed line is the best fit to  conventional AHE theories. The best fit shows significant non-linearity above saturation while the data is linear.}
\label{Fig_2K_conv_Hall_fit}
\end{figure}

Figure \ref{Fig_2K_conv_Hall_fit} shows the Hall resistivity along with the best fit to conventional AHE theories at $T = 2$K. The best fit shows significant non-linearity at high-field while the $\rho_{xy}$ data is linear.

\pagebreak

{
\section{Conventional AHE fit parameters}}

\begin{table}[!h]
\centering
\caption{Conventional anomalous Hall effect fit parameters}
\begin{tabular}{lcccc}
\hline\hline
\multicolumn{1}{c}{$T$ (K)} &  & $R_0$ (m$^3$/C) & $a$ (m/A) & $b$ (V$^{-1}$)\\
\hline
2   &  & {1.43$\times10^{-10}$}    & {-1.59$\times10^{-7}$}     & {0.29}\\
10.5   &  & {-8.23$\times10^{-9}$}    &  {1.20$\times10^{-6}$}    & {-6.40}\\
15   &  & {-1.52$\times10^{-8}$}   & {3.73$\times10^{-6}$} & {-25.1 } \\
\hline \hline
\label{Tab_fit_param}
\end{tabular}
\end{table}

{
Table \ref{Tab_fit_param} shows the parameters obtained from the fit to the conventional anomalous Hall effect equation (Eq. 1 of main text).\\}

{
All parameters vary strongly with temperature. The value of $R_0$ is small and positive at 2K, becomes large and negative and 10.5K and becomes {even larger} at 15K. In contrast, the measured Hall resistivity is linear at high-field at all temperatures with a weakly varying slope {(Section IX)}.

The strong temperature dependence of $a$ and $b$, including sign changes is also unlikely to occur in the conventional theories. These observations are further evidence that the conventional theories cannot satisfactorily explain the behavior of $\rho_{xy}$ in \ce{TmB4}.}

\pagebreak

\section{Comparison of magnetization and transport properties in the two modulated phases}

\begin{figure}[h!]
	\includegraphics[trim=1cm 1cm 1cm 1cm, width=0.6\linewidth]{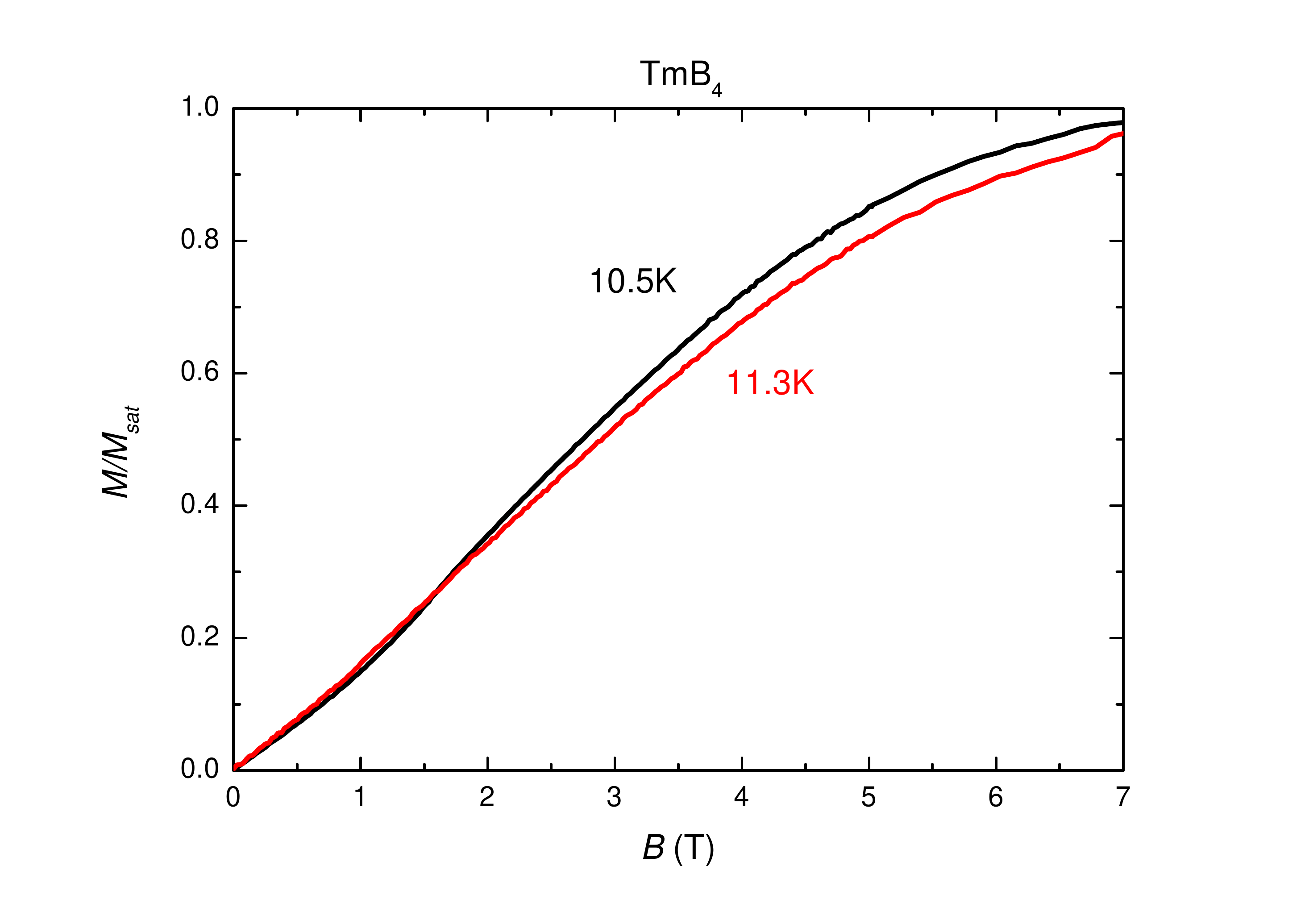}
	\caption{Magnetization of \ce{TmB4} at $T = 10.5$K and $T = 11.3$K. The magnetization curves are qualitatively similar in the two modulated phases. The above curves have not been corrected for demagnetization.}
\label{Fig_mod_phase_magn}
\end{figure}

\begin{figure}[h!]
	\includegraphics[width=\linewidth]{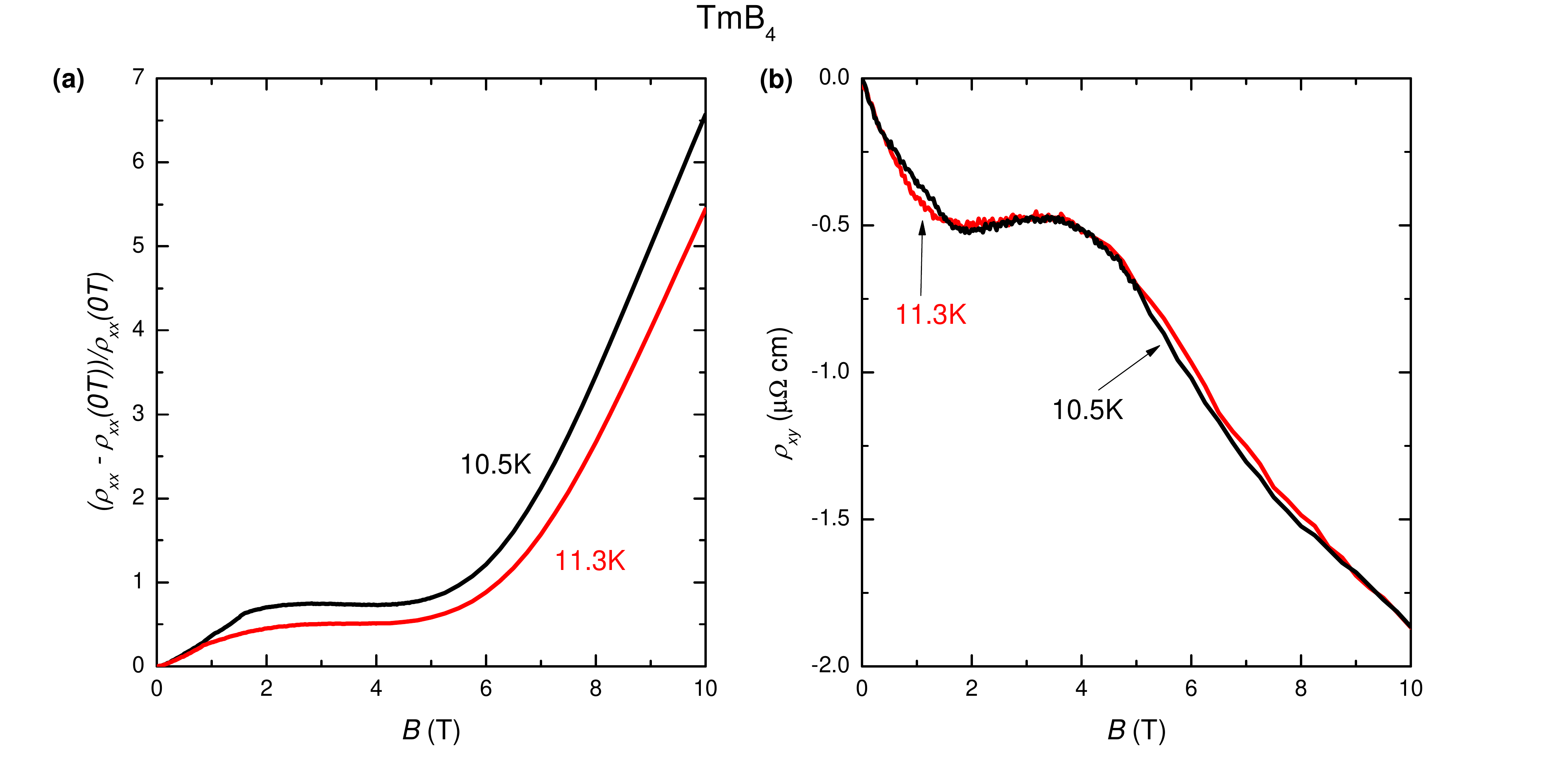}
	\caption{\textbf{(a)} Magnetoresistance of \ce{TmB4} at $T = 10.5$K and $T = 11.3$K. \textbf{(b)} Hall resistivity of \ce{TmB4} at $T = 10.5$K and $T = 11.3$K. The transport properties are qualitatively similar in the two modulated phases. The above curves have not been corrected for demagnetization.}
\label{Fig_mod_phase_transport}
\end{figure}

Figure \ref{Fig_mod_phase_magn} and Figure \ref{Fig_mod_phase_transport} show a comparison of the magnetization and transport properties of \ce{TmB4} in the two modulated phases. $T = 10.5$K corresponds to Modulated Phase II and $T = 11.3$K corresponds to Modulated Phase I. Magnetization, magnetoresistance and Hall resistivity are qualitatively similar, and we chose to focus our analysis in the main text on Modulated Phase II. The same analysis and arguments also apply to Modulated Phase I.

\pagebreak

{\section{Carrier concentration}}

\begin{figure}[h!]
	\centering
	\begin{tabular}{cc}
		\includegraphics[trim=2cm 2cm 2cm 1cm, width=0.45\linewidth]{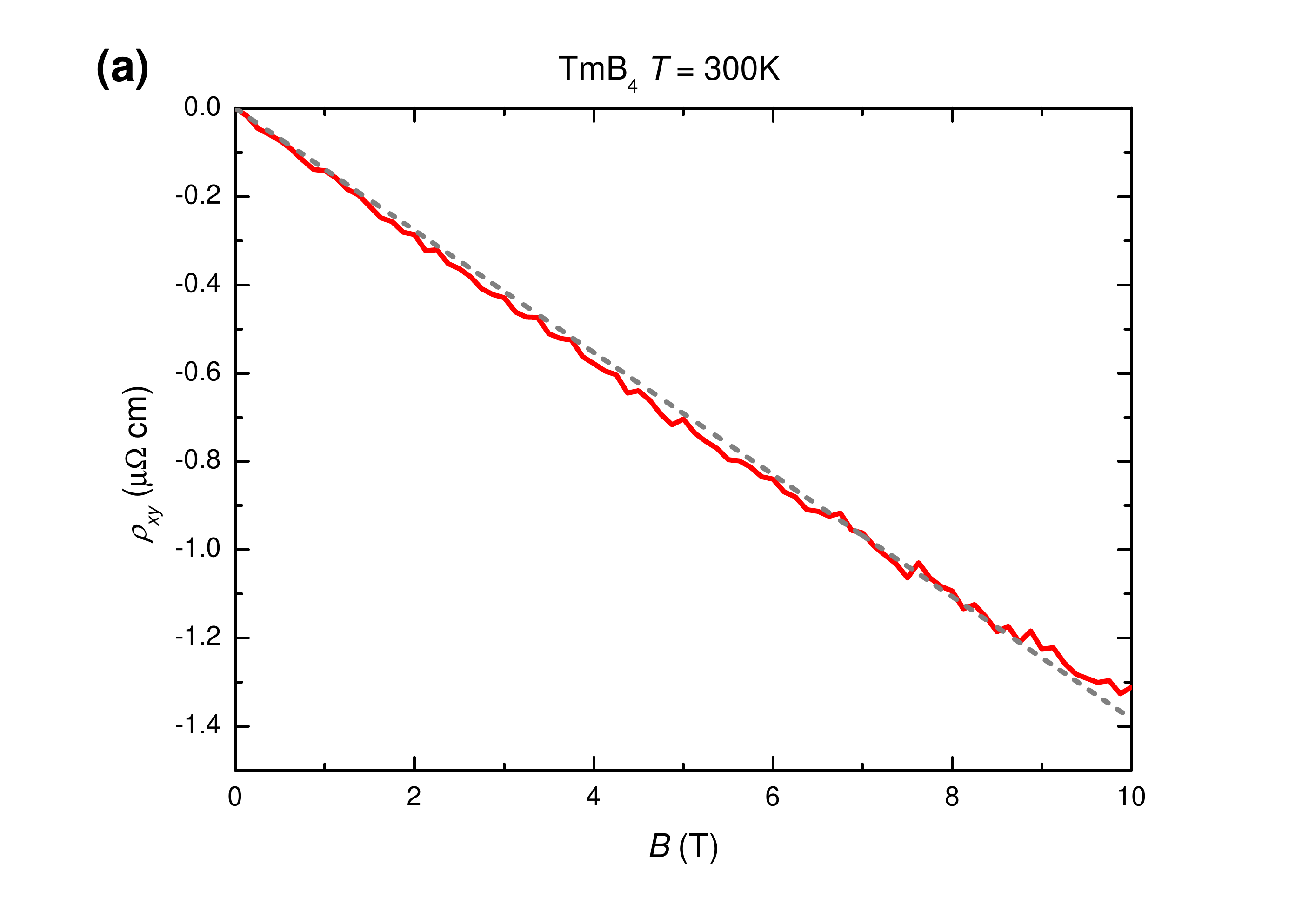} &
		\includegraphics[trim=2cm 2cm 2cm 1cm, width=0.45\linewidth]{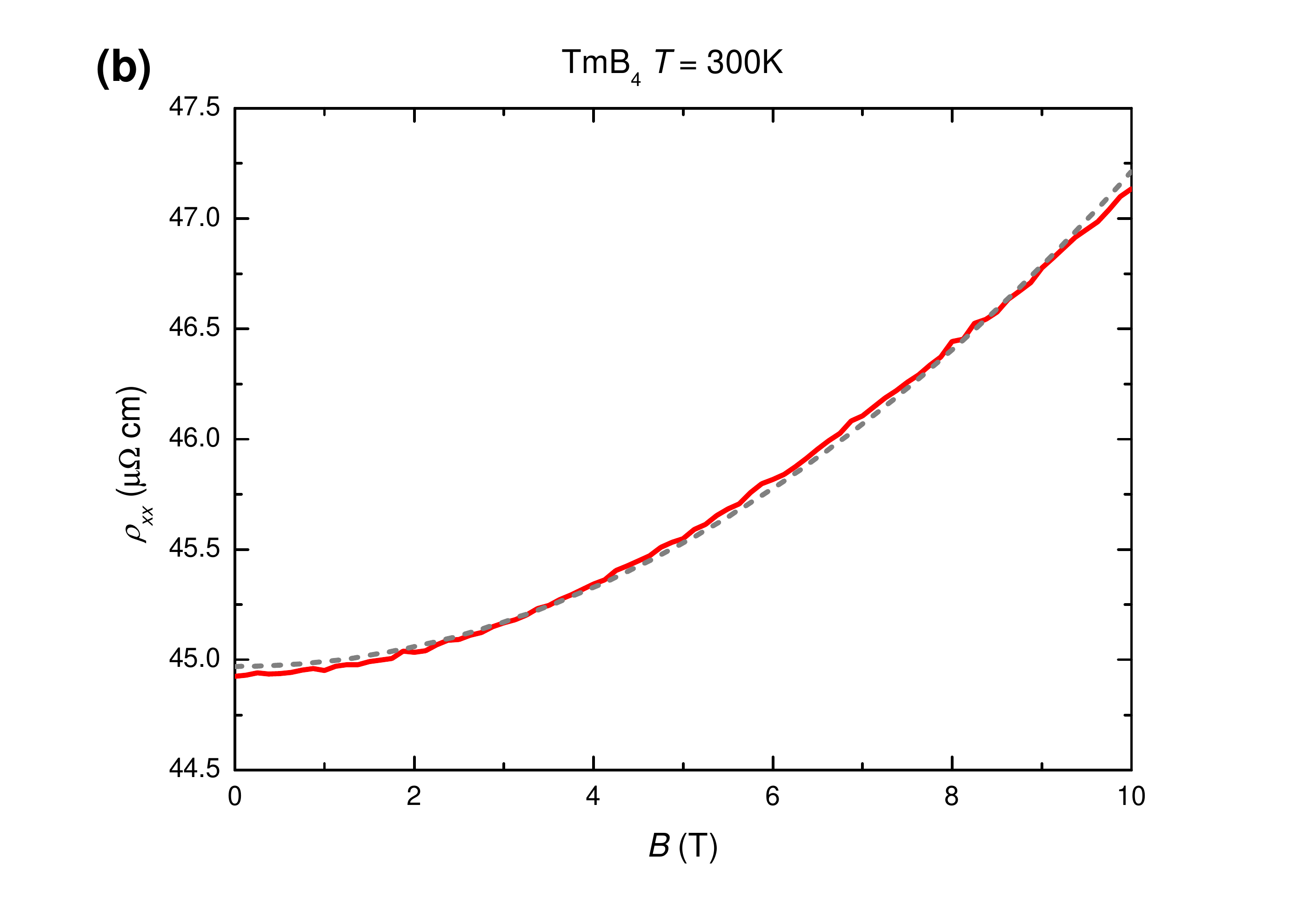}
	\end{tabular}
	\caption{{\textbf{(a)} $\rho_{xy}$ at 300K. A linear fit is shown as a dashed line. \textbf{(b)} $\rho_{xx}$ at 300K. A quadratic fit is shown as a dashed line. The above curves are not corrected for demagnetization, as the magnetization at 300K is very small.}}
\label{Fig_300K_high_field}
\end{figure}

Figure \ref{Fig_300K_high_field} shows the longitudinal and Hall resistivity of \ce{TmB4} at 300K. {We notice that} $\rho_{xy}$ is linear but $\rho_{xx}$ is quadratic and shows no signs of saturation up to the measured field of 10T.
{While a linear field dependence in $\rho_{xy}$ is expected for a single conduction band, a non-saturating magnetoresistance can only appear in two circumstances \cite{Hurd1972}: (1) A Fermi surface with open orbits and (2) two conduction bands with equal and opposite carrier concentration. Ab-initio calculations on \ce{RB4} \cite{Yin2008} and quantum oscillation measurements on the related compound \ce{YB4} \cite{Tanaka1985} show that the Fermi surface consists of multiple closed electron and hole pockets. Therefore, we conclude that the second scenario is occuring in \ce{TmB4}.}\\

{For a two-band model with equal electron and hole concentration, the Hall resistivity and the magnetoresistance are given by \cite{Ashcroft1976}}

\begin{eqnarray}
\rho_{xy} & = & \frac{(\rho_2^2-\rho_1^2)R}{(\rho_1 + \rho_2)^2} {B}, \\
\rho_{xx} & = & { \frac{\rho_1\rho_2}{\rho_1+\rho_2} + \frac{R^2}{\rho_1+\rho_2}{B^2}},
\label{Eqn_MR}
\end{eqnarray}
where $R = 1/ne$, $n$ is the carrier concentration of each band and $\rho_1$ and $\rho_2$ are the resistivities of the two bands respectively. In the above equations $\rho_{xy} \propto B$ and $\rho_{xx}\propto B^2$, exactly as we observe in \ce{TmB4}.\\

{By comparing the parameters obtained from our fits to the 300K data, we obtain $R_{300K} = 2.044 \times10^{-8}$~m$^3$/C, $\rho_{1,300K} = {8.425}\times10^{-7}$~$\Omega$m and $\rho_{2,300K} = {9.615}\times10^{-7}$~$\Omega$m. The carrier concentration at 300K is $n_{300K} = 3.054\times10^{26}$~m$^{-3}$, which corresponds to {0.061} electrons and holes per unit cell.} \\

\begin{figure}[h!]
	\centering
	\begin{tabular}{cc}
		\includegraphics[trim=2cm 2cm 2cm 1cm, width=0.43\linewidth]{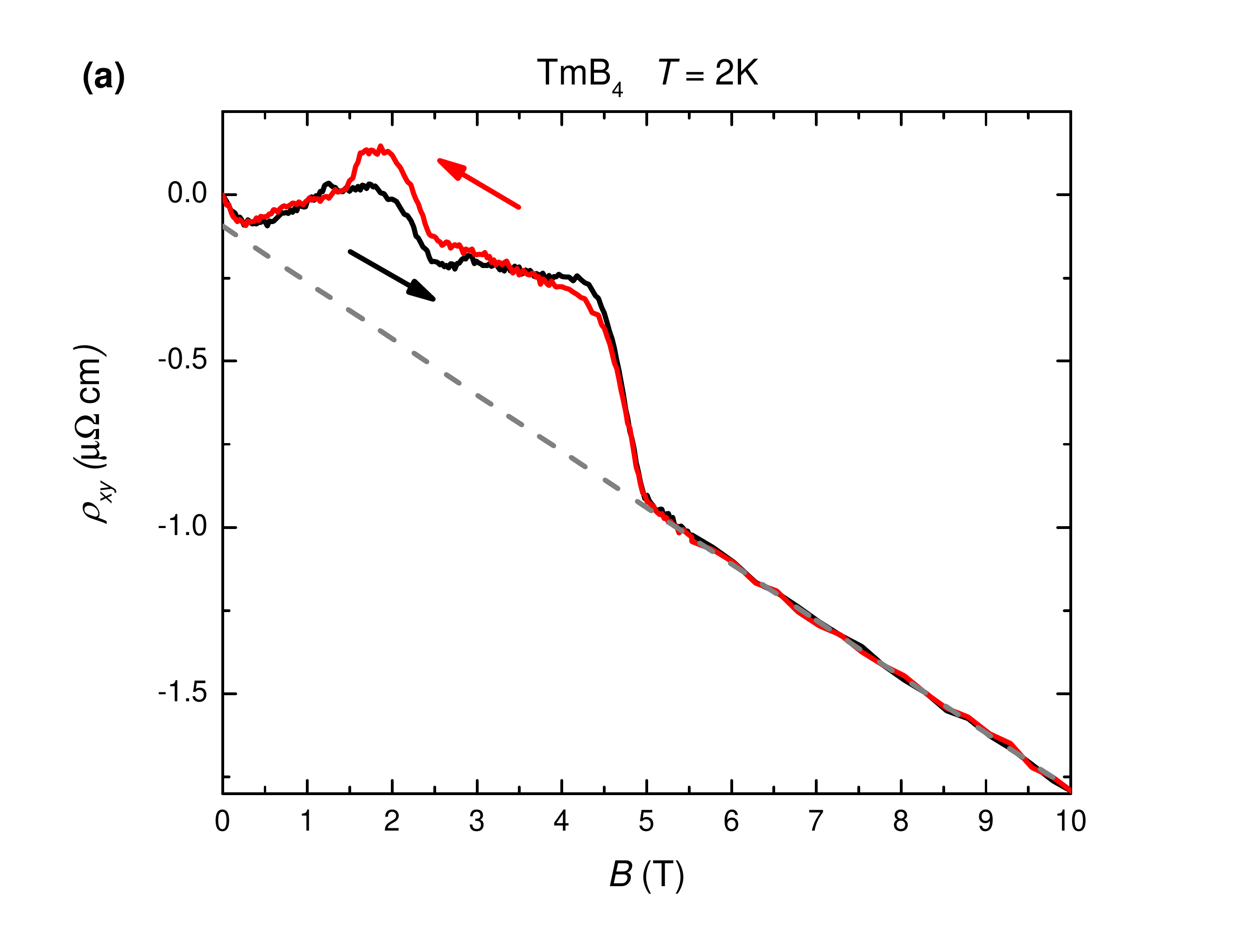} &
		\includegraphics[trim=2cm 2cm 2cm 1cm, width=0.43\linewidth]{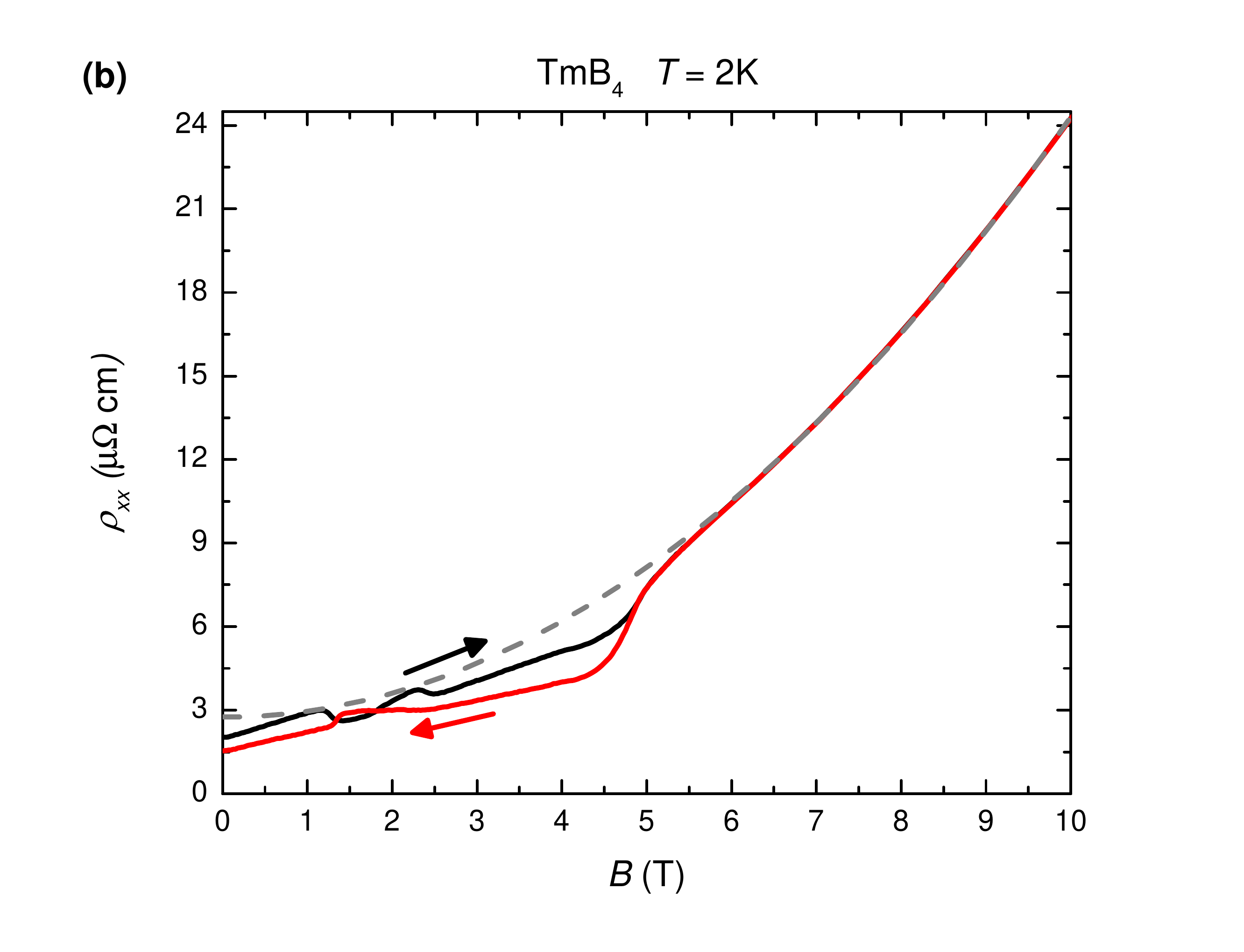}
	\end{tabular}
	\caption{\textbf{(a)} $\rho_{xy}$ at 2K up to 10T. A linear fit to the data above 5T is shown as a dashed line. \textbf{(b)} $\rho_{xx}$ at 2K up to 10T. A quadratic fit to the data between 5T is shown as a dashed line.}
\label{Fig_2K_high_field}
\end{figure}

Figure \ref{Fig_2K_high_field} shows the longitudinal and Hall resistivity of \ce{TmB4} at 2K. {At 2K, $\rho_{xy}$ is linear at high-field but has a non-zero y-intercept. A simple two-band model or the conventional AHE theories cannot account for such a constant term. However, if we only consider the slope of the $\rho_{xy}$ curve, we can repeat the above analysis to obtain the carrier concentration: $R_{2K} = {3.772} \times 10^{-8}$~m$^{3}$/C, $\rho_{1,2K} = {2.042} \times 10^{-8}$~$\Omega$m and $\rho_{2,2K} = {2.255 \times 10^{-8}}$~$\Omega$m. The corresponding carrier concentration is $n_{2K} = {1.655} \times 10^{26}$ m$^{-3}$, which is {0.033} electrons and holes per unit cell.} We note that the carrier concentration at 2K is similar to the value obtained at 300K as well as the experimental value obtained at 1.5K by Tanaka and Ishizawa for the non-magnetic compound \ce{YB4} ($\sim$ 0.03) \cite{Tanaka1985}. {These results allow us to definitively conclude that the high-field behavior of $\rho_{xy}$ at 2K is the sum of the linear contribution from ordinary Hall effect and a constant term.}\\

\begin{figure}[h!]
	\centering
	\begin{tabular}{cc}
		\includegraphics[trim=2cm 2cm 2cm 1cm, width=0.43\linewidth]{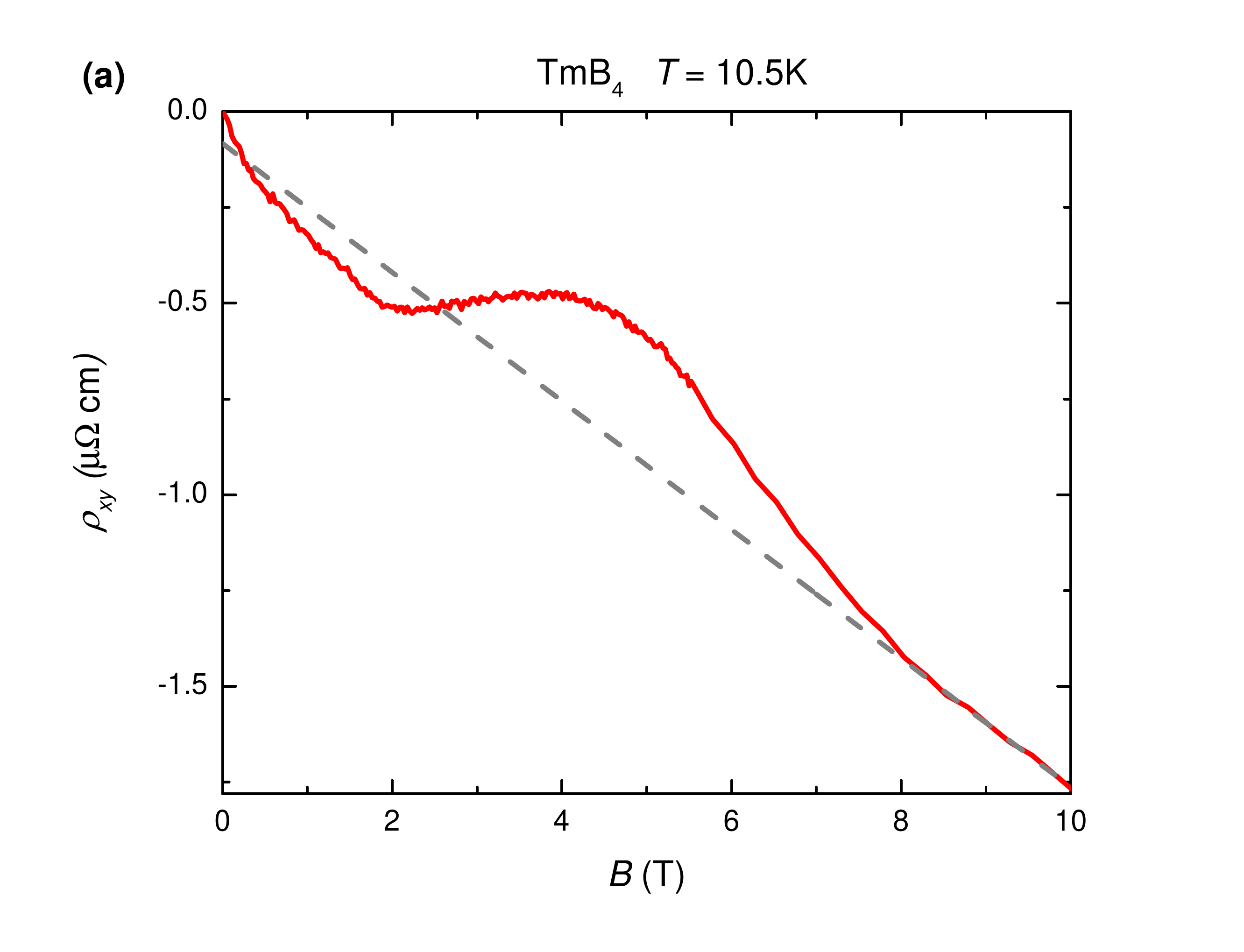} &
		\includegraphics[trim=2cm 2cm 2cm 1cm, width=0.43\linewidth]{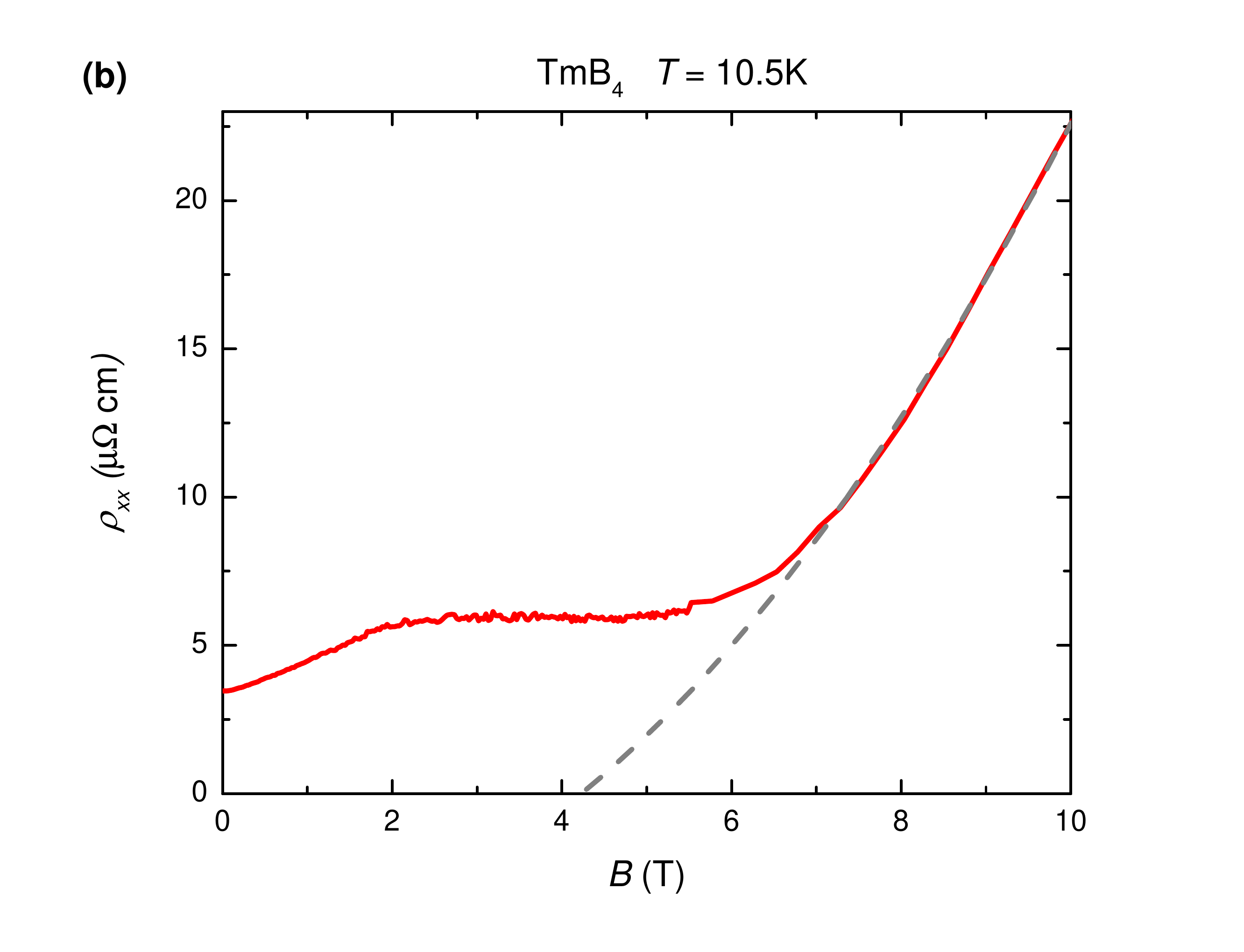}
	\end{tabular}
	\caption{\textbf{(a)} $\rho_{xy}$ at 10.5K. A linear fit to the data above 7T is shown as a dashed line. \textbf{(b)} $\rho_{xx}$ at 10.5K. A quadratic fit to the data above 7T is shown as a dashed line.}
\label{Fig_105K_high_field}
\end{figure}

\begin{figure}[h!]
	\centering
	\begin{tabular}{cc}
		\includegraphics[trim=2cm 2cm 2cm 1cm, width=0.43\linewidth]{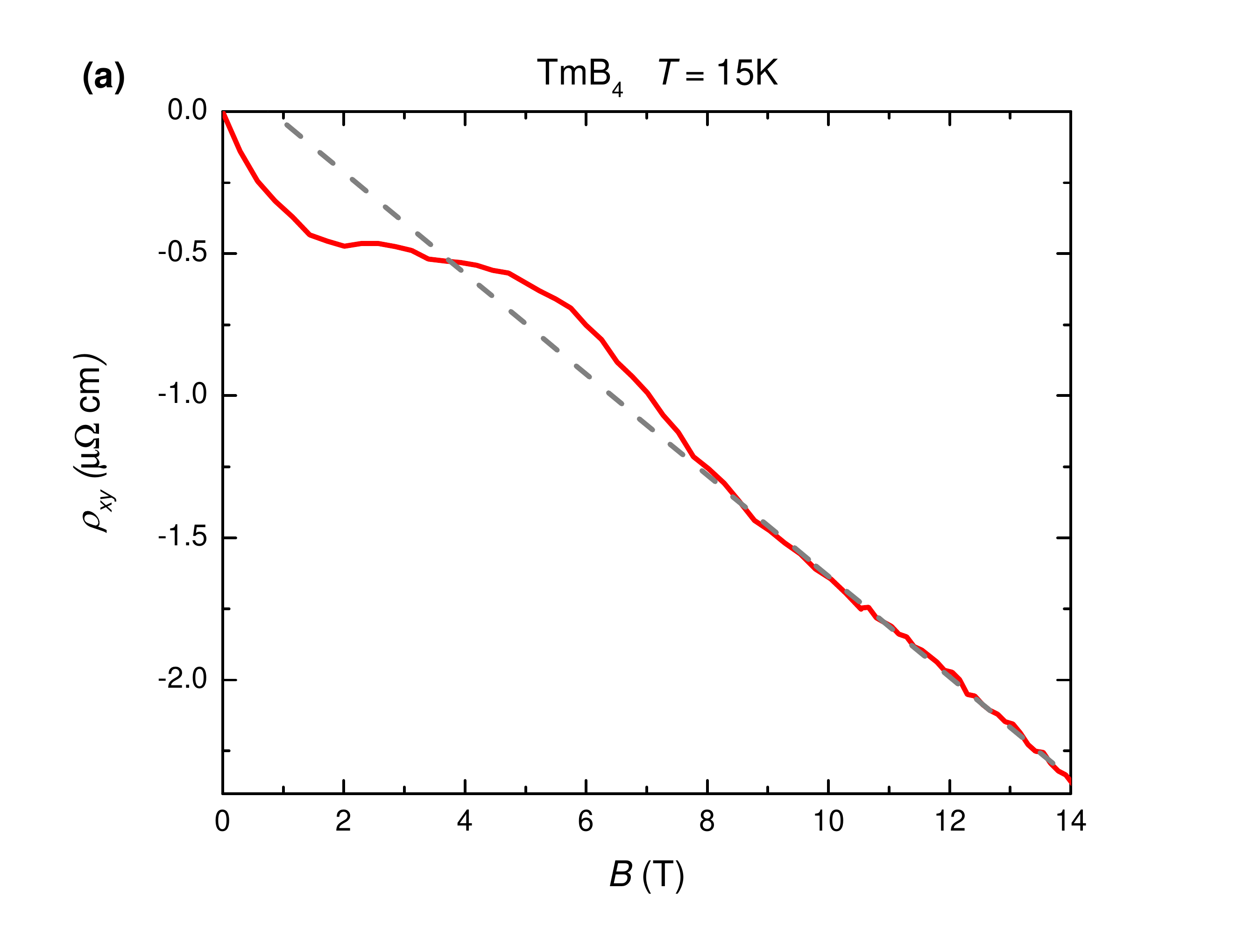} &
		\includegraphics[trim=2cm 2cm 2cm 1cm, width=0.43\linewidth]{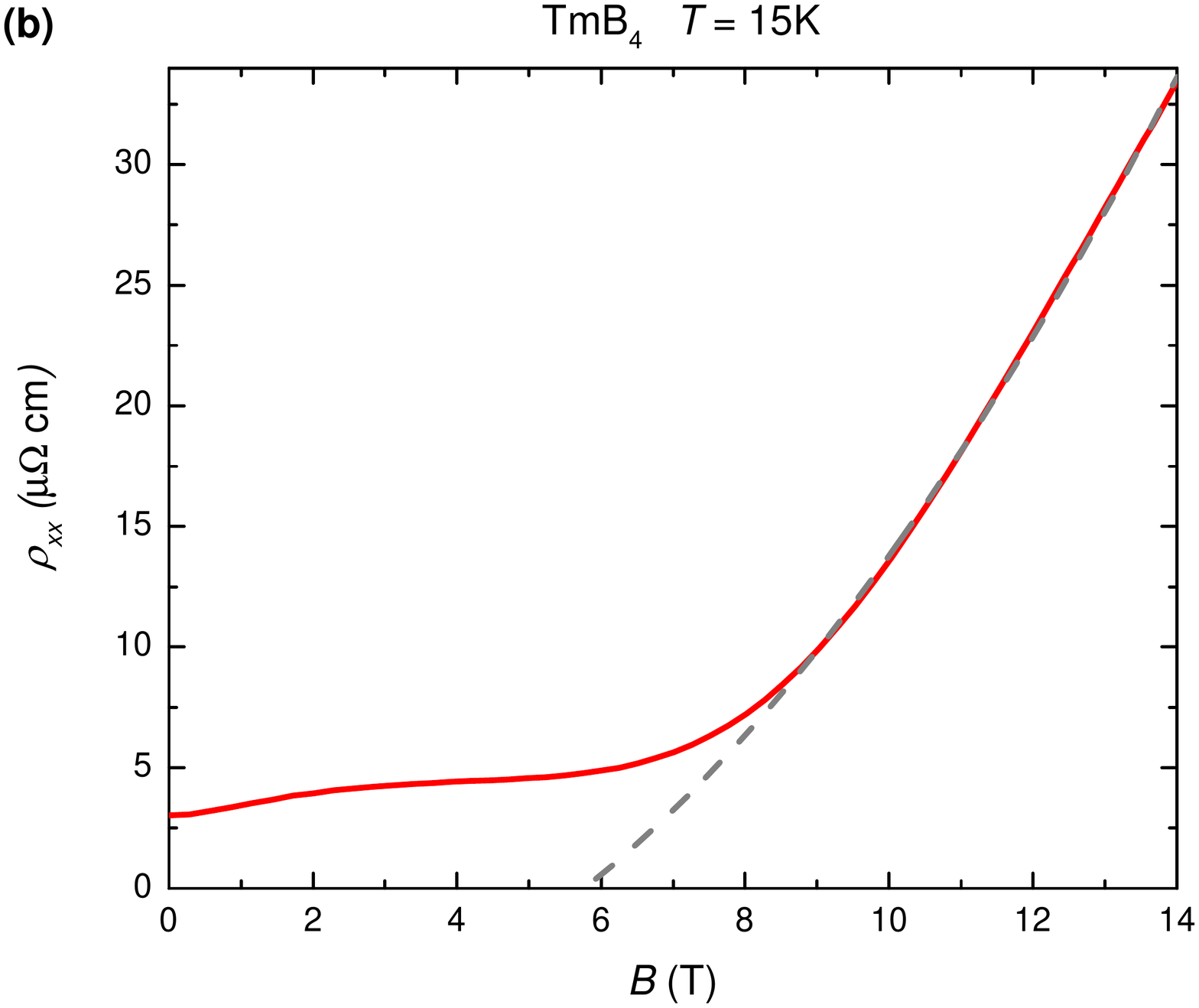}
	\end{tabular}
	\caption{\textbf{(a)} $\rho_{xy}$ at 15K. A linear fit to the data above 8T is shown as a dashed line. \textbf{(b)} $\rho_{xx}$ at 15K. A quadratic fit to the data above 8T is shown as a dashed line.}
\label{Fig_15K_high_field}
\end{figure}

{Figures \ref{Fig_105K_high_field} and \ref{Fig_15K_high_field} show the longitudinal and Hall resistivity of \ce{TmB4} at 10.5K and 15K respectively. The behavior of $\rho_{xy}$ above saturation (Fig. \ref{Fig_105K_high_field}(a) and \ref{Fig_15K_high_field}(a)) is similar to 2K: linear with a non-zero y-intercept. However, attempting to fit the $\rho_{xx}$ data above saturation to Eqn \ref{Eqn_MR} (Fig. \ref{Fig_105K_high_field}(b) and \ref{Fig_15K_high_field}(b)) leads to a negative resistance at zero magnetic field implying that either $\rho_1$ or $\rho_2$ is less than zero. We believe this unphysical result is due to the presence of additional MR contributions of unknown origin at 10.5K and 15K. If we only compare the slopes of the $\rho_{xy}$ curves at 2K, 10.5K and 15K, we see that they are all of a similar value (Table \ref{Tab_slope}). This close correspondence suggests that $\rho_{xy}$ above saturation at 10.5K and 15K is also the sum of the ordinary Hall contribution and a constant term.\\}

\begin{table}[!h]
\centering
\caption{Slope of $\rho_{xy}$ at high field}
\begin{tabular}{lcccc}
\hline\hline
\multicolumn{1}{c}{$T$ (K)} &  & Slope of $\rho_{xy}$ ($\mu\Omega$ cm T$^{-1}$)\\
\hline
2   &  & -0.169   \\
10.5   &  & -0.168    \\
15   &  & -0.178  \\
\hline \hline
\label{Tab_slope}
\end{tabular}
\end{table}

\pagebreak

\bibliographystyle{apsrev}
\bibliography{TmB4-ref}